\documentclass[letters, useAMS, fleqn, usenatbib]{mnras}

\usepackage{amsmath}
\usepackage{amssymb}
\usepackage{booktabs}
\usepackage{xcolor}
\usepackage{graphicx}
\usepackage{dcolumn}
\usepackage{bm}
\usepackage{hyperref}
\usepackage{natbib}
\hypersetup{colorlinks, allcolors={blue}}
\usepackage{soul}
\usepackage{multicol}
\usepackage[normalem]{ulem}
\usepackage{csquotes}
\usepackage{physics}
\usepackage{verbatim}
\usepackage{subcaption}
\usepackage{amsmath}
\usepackage{ulem}

\newcommand{\nb}[1]{\textsf{\color{green}{ #1}}}

\usepackage{orcidlink}
\usepackage{soul}
\usepackage{cancel}

\usepackage[normalem]{ulem}
\usepackage{xcolor}

\title{$f$-mode oscillations in hot Neutron Stars: \\
Effect of hyperons and neutrino trapping}

\author[Barman and Chatterjee]{Nilaksha Barman$^{1}$\,\orcidlink{0009-0008-1220-359X}\thanks{nilaksha.barman@iucaa.in}, 
Debarati Chatterjee$^{1}$\,\orcidlink{0000-0002-0995-2329}\thanks{debarati@iucaa.in}
\\
$^{1}$Inter-University Centre for Astronomy and Astrophysics, Pune University Campus, Pune 411007, India
}

\AtBeginDocument{\RenewCommandCopy\qty\SI}
\begin{document}
\maketitle

\begin{abstract}
In this work, we present an equation of state formalism for hot Neutron Stars (NSs) which consistently includes the effects of finite temperature, hyperons as well as neutrino trapping, relevant for the study of proto-neutron stars, binary neutron star mergers and supernova explosions. Within a non-linear relativistic mean field description, the framework allows for a systematic variation of nuclear parameters within the range allowed by uncertainties in nuclear experimental data, ensuring compatibility with nuclear theory, astrophysical and heavy-ion data. We then investigate the role of nuclear and hypernuclear parameters as well as thermal effects on NS macroscopic properties and $f$-mode oscillations in hot neutron stars within Cowling approximation. Our results reveal that in hyperonic neutron stars with trapped neutrinos, the saturation nuclear density shows moderate to strong correlation with NS astrophysical observables. We also investigate whether thermal effects break universal relations and provide fit relations for hot NS configurations in the neutrino-trapped regime.
\end{abstract}


\section{Introduction}\label{sec:intro}

The matter in mature neutron stars (NSs) is well described as cold (that is at temperatures well below the Fermi temperature) and in $\beta$ equilibrium under the condition of charge neutrality.
The thermodynamic condition for such a system is described by the microscopic Equation of State (EOS), which is a relation among thermodynamic variables of the system. The pressure-energy density relation provides information about macroscopic structure of NSs through the Tolman-Oppenheimer-Volkoff (TOV) equation. The EOS is highly uncertain because of intrinsic uncertainty in the nuclear physics properties, i.e., the so-called nuclear saturation properties 
~\citep{Glendenning,Schaffner-Bielich_2020,Lattimer_Prakash_2004science,Lattimer_2012annurev-nucl,Haensel_Potekhin_2007,Baym_Hatsuda_2018rpp,Burgio_Schulze_2021ppnp,Burgio_Fantina_2018assl,Blaschke_Chamel_2018assl}.
Further at higher density neutron chemical potential becomes large enough to allow for appearance of non-nucleonic degrees of freedom (d. o. f.) such as hyperons, meson condensates, delta baryons, deconfined quarks etc.
The difficulty in reconciling the softening of EOS  due to hyperons with observation of NS with masses around $2M_{\odot}$~\citep{Demorest_Pennucci_2010nature,Antoniadis_2013science} gave rise to the so called hyperon puzzle~\citep{Lonardoni_2015}.
However, possible solutions to the problem can be achieved via  sufficiently repulsive hyperon-nucleon and hyperon-hyperon interactions, which in turn provide information regarding the hyperonic EOS (hereafter denoted by NY) and hypernuclear potentials (e.g. see discussions in~~\citep{Chatterjee_Vidana_epja2016,Bombaci_2017,Oertel_Hempel_2017revmodphys,Logoteta_2021,Vidana_2022} and references therein). \\


Neutron stars are well known to emit electromagnetic radiation in radio, X-ray, and gamma-ray frequencies; in addition, they can emit gravitational waves, offering a distinct and complementary channel of observation. Although NS masses can be determined with high precision in binary systems through the measurement of post-Keplerian parameters, the determination of NS radius from thermal emission studies suffers from large uncertainties~\citep{Lattimer_2021ann_rev}.
The Neutron Star Interior Composition Explorer (NICER)~\citep{NICER_2014} onboard the International Space Station (ISS) has recently provided improved measurements of masses and radii of galactic pulsars by the pulse profile modeling of hotspots. EM data from NICER and other telescopes such as XMM-Newton impose mass-radius constraints on some known pulsars: PSR J0030-0451 \citep{Miller_2019,Riley_2019}, PSR J0740-6620~\citep{Miller_2021,Riley_2021} and PSR J0437-4715~\citep{Rutherford_Mendes_2024apjl}. 
PSR J0740-6620 is currently recognized as the most massive pulsar discovered, establishing a lower bound on the Tolman–Oppenheimer–Volkoff (TOV) limit for neutron stars.
It has a gravitational mass of $2.08^{+0.07}_{-0.07}$M$_{\odot}$~\citep{Miller_2021,Riley_2021}. 
On the other hand, the analysis of the inspiral data from only confirmed multi-messenger observation from the Binary Neutron Star (BNS) merger Gravitational Wave (GW) event GW170817~\citep{LSC_2017_GW170817} 
provides joint posteriors on macroscopic properties of the binary components like masses, radii and tidal deformabilities~\citep{LSC_2018_GW170817,LSC_2019_GW170817}, which in turn provides information about the internal structure and composition. More BNS events are expected with future GW observations and 3rd generation detectors such as the Cosmic Explorer (CE) and Einstein Telescope (ET), which will be sensitive to merger-ringdown phases of the GW signals. \\


In astrophysical scenarios such as core collapse supernovae (CCSN)~\citep{Mezzacappa_2015arxiv,Janka_Langanke_2007physrep,O’Connor_Couch_2018apj,Burrows_Vartanyan_2020nature,Murphy_Ott_2009apj,Radice_Morozova_2019apjl}, Proto-Neutron Stars (PNSs)~\citep{Pons_Reddy_1999apj} and binary Neutron Star (BNS) mergers~\citep{Rosswog_2015ijmpd,Baiotti_Rezzolla_2017rpp,Endrizzi_Logoteta_2018prd,Baiotti_2019ppnp,Most_Motornenko_2023prd} nuclear densities of about 0.5 - 6 times the nuclear saturation density, temperatures as high as 100 MeV and the charge fraction in the range of 0.01 - 0.60~\citep{Oertel_Hempel_2017revmodphys} can be reached. Under these conditions, it becomes thermodynamically favorable for non-nucleonic d. o. f. to appear in the core. They may even appear at lower baryon densities compared to the $T=0$ case~\citep{Fortin_Oertel_2018pasa,Stone_Dexheimer_2021mnras,Raduta_Oertel_2020mnras,Sedrakian_Harutyunyan_2021universe,Tsiopelas_Sedrakian_2024}. 
These scenarios require a consistent treatment of finite temperature and out-of-$\beta$-equilibrium effects. 
A finite temperature EOS (also called General Purpose EOS) is completely specified by three conditions: the equilibrium temperature ($T$), baryon density ($n_B$) and charge fraction ($Y_Q$). For studies of finite temperature astrophysical scenarios in numerical simulations, a wide range and finely spaced sets of these EOS parameters are required for NS matter with different composition. 
In the absence of EOS models spanning the required parameter ranges and composition and due to the complexity of using tabulated realistic EOSs, the so-called $\Gamma$-law (supplementing the cold EoS with ideal gas thermal contributions) has often been used in simulations. However, this approach to describe thermal contributions in NSs has recently been shown to be inappropriate, particularly for matter consisting of non-nucleonic degrees of freedom \citep{Raduta_2022epja,Kochankovski_Ramos_2022mnras, Blacker_Kochankovski_2024prd}. 


 A limited number of General Purpose EOSs are available in the online CompOSE~\citep{CompOSE} database which have been employed recently in numerical simulations~\citep{Blacker_Bauswein_2023prd,Blacker_Kochankovski_2024prd}. These include Single Nucleus Approximation (SNA) EOSs: LS220~\citep{Lattimer_Swesty_1991npa} and STOS~\citep{Shen_Toki_1998npa}. There are also EOSs which include purely nucleonic or nucleon + other d. o. f. These models are based on Density Dependent Relativistic Mean Field (DD-RMF)~\citep{Typel_Wolter_1999npa}, Non-Linear Relativistic Mean Field (NL-RMF)~\citep{Chen_Piekarewicz_2014prc} and Chiral Mean Field (CMF)~\citep{Dexheimer_Schramm_2008apj} theory. 
Available general purpose EOSs with hyperons based on these formalisms include~\cite{Banik_Hempel_2014apj_supplement},~\cite{Marques_Oertel_2017prc},~\cite{Dexheimer_Schramm_2008apj},~\cite{Fortin_Oertel_2018pasa},~\cite{Gulminelli_Raduta_2013prc},~\cite{Ishizuka_Ohnishi_2008jpg},~\cite{Kochankovski_Ramos_2022mnras},~\cite{Stone_Dexheimer_2021mnras},~\cite{Shen_Toki_2011apj},~\cite{Tsiopelas_Sedrakian_2024},~\cite{Raduta_2022epja}.
One must note here that these are three-dimensional tabulated EOSs and each correspond to only one set of nuclear parameterizations among other possible ones within the allowed uncertainty in nuclear experimental observables. Hence they do no allow systematic variation of these nuclear parameters to probe their role in controlling global NS observables. There exist only a few recent works that have varied some of the nuclear parameters (see e.g.~\cite{Tsiopelas_Sedrakian_2024}), but their correlation with other nuclear parameters or with observable properties of hot NSs is not evident from such studies. Further, at high temperature and density the neutrino mean free path can be sufficiently smaller
than the size of the star, allowing them to be in equilibrium with the thermal bath. 
Studies on BNS mergers~\citep{Alford_2018,Alford_2021} have indicated that at high temperatures (above $\sim 5$ MeV) neutrinos may be trapped, depending on the neutrino mean free path at a given density.
The role of these trapped neutrinos has also been shown to be important in the case of PNS evolution~\citep{Pons_Reddy_1999apj,Raduta_Oertel_2020mnras}. In case of trapped neutrinos, the thermodynamic system is specified by the total lepton fraction instead of charge/electron fraction.
We note that the general purpose EOSs available in CompOSE do not include the neutrino component, which is added in simulations as a separate component. However the comparison between neutrino-free and neutrino-trapped matter can provide important insight into their role on oscillation modes as well as other properties in hot NS systems. 


Finite temperature effects manifests itself in macroscopic properties as well as asteroseismology of hot and dense NS environment. Several recent works have investigated thermal effects by solving for nonradial oscillations of spherically
symmetric NSs within the Cowling approximation~\citep{Thapa_Beznogov_2023prd} and full general relativity~\citep{Barman_Pradhan_2025prd}. Hydrodynamical simulations have also been carried out to study the effect of various eigenmodes ($f-$, $p-$, $g-$modes) and how they impact GW emission~\citep{Torres2018,Torres2019,Sotani2020c,Rodriguez2023,Sotani2024a,Sotani2016,Sotani:2019b,Sotani2020,Afle_Kundu_2023prd}. \cite{Sotani2020b} found that $g_1$ and $f$-modes are the dominant frequency modes emitted during the earlier and later stages of PNS evolution. Subsequently, the prospect of inferring PNS properties from asteroseismology through GW data has been explored~\citep{Bruel2023,Bizouard2020}. The effects of finite temperature are also relevant in the still undetected post-merger GW signal of BNS mergers, as well as for studying stability of post-merger remnants. It has been shown in previous studies that appearance of hyperons can alter the Gravitational Waves emitted from the post-merger hypermassive neutron star (HMNS) remnant~\citep{Sekiguchi_Kiuchi_2011prl,Radice_Bernuzzi_2017apjl}. The effects of finite temperature in case of BNS merger scenario with hyperonic matter has been studied recently~\citep{Blacker_Kochankovski_2024prd,Kochankovski_Lioutas_2025prd}. These studies indicate that BNS post-merger properties such as peak frequency and threshold mass are strongly dependent on the underlying hot nuclear EOS and a simple $\Gamma$-law substitution is not sufficient for its description (especially for non-nucleonic d. o. f.). 
Recent attempts have been made at exploring the effect of neutrinos in the thermal bath with nucleonic and non-nucleonic d. o. f.~\citep{Malfatti_Orsaria_2019prc,Raduta_Oertel_2020mnras,Issifu_Marquez_2023mnras,Laskos-Patkos_Pavlos_2023hnpsanp,Ghosh_Shaikh_2024npb,Kumar_Thakur_2024mnras,da_Silva_Issifu_2025arxiv,Issifu_da_Silva_2024epjc,Issifu_Menezes_2025jcap,Zheng_Sun_2025arxiv,Thakur_Issifu_2025arxiv}.
These studies however consider selected EOS parametrization and do not take into account the large range of nuclear and other uncertainties that are allowed in their respective parameter space. 


In a series of publications~\citep{Ghosh_Chatterjee_2022epja,Ghosh_Pradhan_2022_fspas,Shirke_Ghosh_2023apj}, we developed a formalism  within a non-linear relativistic mean field model (NL-RMF) framework to constrain the parameter space of cold NS EOS using information from multi-disciplinary physics, i.e., microscopic calculations from Chiral Effective Field Theory ($\chi$EFT) at low density (0.5 - 1.4~$n_{sat}$)~\citep{Drischler_Carbone_2016prc,Drischler_Hebeler_2019prl} and astrophysical observations (Astro) at high density (beyond 3~$n_{sat}$). At intermediate density (1-3 $n_{sat}$) we also have constraints from experimental data from Heavy Ion Collisions (HIC) (see detailed discussions on the constraints in Sec.\ref{subsec:constraints}. These works probed possible correlations between the nuclear  parameters with NS global properties (e.g. radius, tidal deformability) and found the in-medium nucleon mass $m^*$ to be the most strongly correlated to these observables. Other studies~\citep{Jaiswal2021,Pradhan_Chatterjee_2021prc,Pradhan_Chatterjee_2022prc} also showed that $m^*/m$ is also the most dominant parameter that governs fundamental ($f$-) mode characteristics. In order to investigate whether such correlations also hold in finite temperature scenarios, such as oscillations in newly born NSs or binary NS mergers, we extended this Bayesian analysis to include thermal effects ensuring thermodynamic consistency in our previous work~\citep{Barman_Pradhan_2025prd} for nucleonic matter. The role of nuclear saturation parameters as well as thermal properties i.e., entropy per baryon, $S/A$ and out-of-$\beta$ equilibrium condition (charge/electron fraction, $Y_Q$) on the EOS, macroscopic properties as well as oscillations modes in hot and dense matter were investigated. Using the Bayesian posteriors, we searched for possible correlations between nuclear saturation parameters as well as thermal properties with astrophysical observables. We also investigated the effect of nuclear saturation and thermal properties on the Universal relations. However in this work, we did not include the effects of neutrino trapping, which could become relevant at high temperatures. We had also neglected non-nucleonic degrees of freedom such as hyperons, which are expected to appear at the high densities. 

In this work we go beyond our previous study to include hyperonic degrees of freedom as well as the effects of neutrinos in equilibrium with the thermal bath, relevant in astrophysical scenarios at high densities and temperatures, such as NS mergers and proto-neutron stars. 
We use the same NL-RMF framework to describe the EOS, also including hyperonic degrees of freedom. We take into account the same uncertainty range in nuclear properties constrained using multi-disciplinary information as in our previous work and also include uncertainties in hypernuclear properties following~\cite{Ghosh_Pradhan_2022_fspas}. In order to demonstrate the thermal effects, we first consider a fixed nuclear parameter set with different thermodynamic conditions, as in our previous work~\citep{Barman_Pradhan_2025prd}. We then compare between $\nu$-free and $\nu$-trapped hot NS configurations. We explore how the presence of hyperons and trapped neutrinos affect the NS configurations and $f$-mode oscillations. We also examine the $f$-mode frequency relation with average mass density and $C-Love$ Universal Relation under these conditions. 


The paper is structured as follows: in Section~\ref{sec:formalism}, we describe the Non-Linear Relativistic Mean Field formalism followed by a brief overview of different thermodynamic ensembles relevant in the study (including neutrino trapping). In Section~\ref{subsec:gamma_law}, we define the so-called $\Gamma$-law used in the study of finite temperature EOSs. Then in Section~\ref{subsec:constraints} we 
describe the Bayesian scheme employed in this study, recapitulate the multi-physics constraints and provide the table for nuclear and hypernuclear parameters. In Section~\ref{sec:results}, we first investigate the thermal effects on N- and NY-matter considering a fixed parameterization of nuclear and hypernuclear parameters. 
We first compare isothermal configurations for two select thermodynamic states in $\nu$-free matter (Sec.~\ref{subsec:nu_free}). In Sec.~\ref{subsec:nu_free_vs_trapped}, we then consider isentropic hot NS configurations in neutrino-free as well as neutrino-trapped regimes for N- and NY-matter. 
In Sec.~\ref{subsec:PNS_constraints}, we then allow for a systematic variation of nuclear and hypernuclear parameters as well as thermal effects to perform a full Bayesian study. We consider isentropic configurations in $\nu$-trapped matter for two thermal configurations to study their macroscopic properties and $f$-mode oscillations. The $f-$modes are calculated within the relativistic Cowling approximation~\citep{Cowling}. In this section, we also identify the dominant nuclear parameters that control the hot NS properties. This is followed by a discussion on relation between $f$-mode frequency and average mass density ($\nu_f-\sqrt{\bar{M}/\bar{R}^3}$) and $C-Love$ relations for these NS configurations. In Sec.~\ref{sec:discussions} we provide a summary of our results as well as a comparison of the results with other works. In the entire analysis, we have used natural units ($c=\hbar=k_B=G=1$).

\section{Formalism}
\label{sec:formalism}

We use a relativistic mean field (RMF) model with non-linear coupling~\citep{Chen_Piekarewicz_2014prc,Hornick_Tolos_2018prc} to describe homogeneous nuclear matter inside hot neutron stars. The baryon interactions are described via the exchange of $\sigma$, $\omega$ and $\Vec{\rho}$ mesons. 
The $\sigma$ (scalar) and $\omega$ (vector) mesons are responsible for the long-range attractive and short-range repulsive components of the baryon–baryon interaction respectively, while
the $\Vec{\rho}$ vector meson is responsible for isospin asymmetry. 
For hyperons, the strange vector meson $\phi$ describes hyperon-hyperon repulsion. We ignore the strange scalar meson $\sigma^*$ in this work as it softens the EOS making it incompatible with observed NS masses (see~\cite{Weissenborn_Chatterjee_2012npa}). \\

The Lagrangian density describing the baryon-baryon interaction in the non-linear RMF model~\citep{Chen_Piekarewicz_2014prc,Hornick_Tolos_2018prc} is given by
\begin{align} \label{eq:NL_RMF}
    \mathcal{L} &= \sum_{B}\Bar{\psi}_B (i \gamma_{\mu} \partial^{\mu} - m_B + g_{\sigma B} \sigma - g_{\omega B} \gamma^{\mu} \omega_{\mu} - g_{\phi B} \gamma^{\mu} \phi_{\mu} \nonumber \\
    &- g_{\rho B} \gamma^{\mu} \rho_{\mu}^a \tau^a )  \psi_B \nonumber \\
    &+ \frac{1}{2} \partial^{\mu} \sigma \partial_{\mu} \sigma - \frac{1}{2} m_{\sigma}^2 \sigma^2 - \frac{1}{4} \omega_{\mu \nu} \omega^{\mu \nu} + \frac{1}{2} m_{\omega}^2 \omega^2 \nonumber \\
    &- \frac{1}{2} \Vec{\rho}_{\mu \nu} \Vec{\rho}^{\mu \nu} + \frac{1}{2} m_{\rho}^2 \Vec{\rho}^2 \nonumber \\
    &- \frac{1}{4} \phi_{\mu \nu} \phi^{\mu \nu} + \frac{1}{2} m_{\phi}^2 \phi^2 \\
    &- U_{\sigma} + \Lambda_{\omega} (g_{\rho N}^2 \Vec{\rho}_{\mu}.\Vec{\rho}^{\mu}) (g_{\omega N}^2 \omega_{\mu} \omega^{\mu}) \nonumber \\
    &+\sum_{L} \Bar{\psi}_L (i \gamma_{\mu} \partial^{\mu} - m_L) \psi_L \nonumber \\
    &+\sum_{i = B,L}\Bar{\psi}_i(-q_i \gamma_{\mu} A^{\mu})\psi_i -\frac{1}{4}F_{\mu\nu}F^{\mu\nu}~ \nonumber~,
\end{align}
 
Here $B$ stands for baryon octet [n, p, $\Lambda$, $\Sigma^-$, $\Sigma^0$, $\Sigma^+$, $\Xi^-$, $\Xi^0$].
The first line in the Lagrangian density represents baryonic contribution, where $\psi_B$ is the Dirac spinor for the $B^{th}$ baryonic species with mass $m_B$ and $\Vec{\tau}$ is the isospin operator. 
The second, third and fourth lines describe free mesonic contributions to the Lagrangian density and $m_{\sigma}$, $m_{\omega}$, $m_{\rho}$ and $m_{\phi}$ are masses of $\sigma$, $\omega$, $\Vec{\rho}$ and $\phi$ mesons respectively. Also $\omega_{\mu \nu}$, $\Vec{\rho}_{\mu \nu}$, $\phi_{\mu \nu}$ are defined as,
\begin{align}
    \omega_{\mu \nu} &= \partial_{\mu} \omega_{\nu} - \partial_{\nu} \omega_{\mu} \\
    \Vec{\rho}_{\mu \nu} &= \partial_{\mu} \Vec{\rho}_{\nu} - \partial_{\nu} \Vec{\rho}_{\mu} \\
    \phi_{\mu \nu} &= \partial_{\mu} \phi_{\nu} - \partial_{\nu} \phi_{\mu}~.
\end{align}
The fifth line in equation~\ref{eq:NL_RMF} includes the $\sigma$-meson self-interaction term
$U_{\sigma} = \frac{1}{3}b m_N (g_{\sigma N} \sigma)^3 + \frac{1}{4}c (g_{\sigma N} \sigma)^4$
and a term describing the non-linear interaction between $\Vec{\rho}$ and $\omega$ meson fields. 
The Lagrangian density also includes the leptonic contribution, where $\psi_L$ is the Dirac spinor for the $L^{th}$ leptonic species (here $L$ represents electron-type leptons [$e^-$, $\nu_e$]) with mass $m_L$. The last line describes electromagnetic interaction through the electromagnetic field $A_{\mu}$ and the second term here is the free field term of the electromagnetic field. $F_{\mu \nu}$ is defined as,
\begin{equation}
    F_{\mu \nu} = \partial_{\mu} A_{\nu} - \partial_{\nu} A_{\mu}~.
\end{equation}
The electric charge of the $i^{th}$ baryonic or leptonic species is given by $q_i$. 
The isoscalar baryon-meson couplings ($g_{\sigma N},g_{\omega N}, b, c$) are fitted to reproduce isoscalar saturation properties (saturation density $n_{sat}$, energy per particle $E_{sat}$, compressibility $K_{sat}$, effective nucleon mass $m^*$) while isovector coupling constants ($g_{\rho N},~\Lambda_{\omega}$) 
 are fitted to reproduce isovector saturation properties (symmetry energy $J_{sym}$ and its slope $L_{sym}$). The isoscalar hyperonic couplings can be fitted to obtain hypernuclear potentials ($U^N_{\Lambda}$, $U^N_{\Sigma}$ and $U^N_{\Xi}$). The hyperonic couplings to the meson fields are related to the hypernuclear potentials by the following relation,
 \begin{equation}\label{eq:hyp_coup}
     U^N_i = -g_{\sigma i} \Bar{\sigma}_0 + g_{\omega i} \Bar{\omega}_0~.
 \end{equation}
 The bar above meson fields represent their mean field values. The $\omega$- and $\phi$-couplings are fixed to their SU(6) values~\citep{Schaffner_Carl_1993prl,Weissenborn_Chatterjee_2012npa}. Hence the uncertainty in the hypernuclear potentials are reflected as uncertainty in the $\sigma$-Y couplings.
 With this prescription in hand, we can write the entropy density of the homogeneous matter as,

 \begin{align}\label{eq:entropy_den}
    s &= -\sum_{i = B,L} \frac{2 J_i+1}{2\pi^2} \int dk k^2 \left[f_{FD}\left(\frac{E_i(k) - \mu_i^*}{T}\right)\right. \nonumber \\
    &\left .\ln f_{FD}\left(\frac{E_i(k) - \mu_i^*}{T}\right) + \Bar{f}_{FD}\left(\frac{E_i(k) - \mu_i^*}{T}\right)\right. \nonumber \\
    &\left . \ln \Bar{f}_{FD}\left(\frac{E_i(k) - \mu_i^*}{T}\right)\right] \nonumber \\
    &+ (\mu^*_i\rightarrow-\mu^*_i) + s_{\gamma}~,
\end{align}

In equation~\ref{eq:entropy_den}, $f_{FD}(x)$ is the Fermi-Dirac distribution and $T$ is the temperature of the local thermal equilibrium. On the other hand, $\Bar{f}_{FD}(x) = 1 - f_{FD}(x)$ is the Fermi-Dirac distribution for non-occupancy of a state. $J_i$ is the spin of the i-th baryonic/leptonic species. 
$\mu_i^*$ represents the effective chemical potential of the $i^{th}$ baryonic/leptonic species. 
For anti-particle contributions from Fermionic species, signs of effective chemical potentials in the integrands of equation~\ref{eq:entropy_den} have to be reversed and this is indicated by the term $(\mu^*_i\rightarrow-\mu^*_i)$. 
The effective chemical potential is related to the actual chemical potentials $\mu_B$ (
baryon chemical potential), $\mu_Q$ (
charge chemical potential) and $\mu_L$ (
lepton chemical potential) by the following relation,
\begin{equation}\label{eq:chem_pot}
    \mu_i^* = B_i \mu_B + Q_i \mu_Q + L_i \mu_L - g_{\omega i} \Bar{\omega}_0 - g_{\rho i} \Bar{\rho}_{03} I_{3i} / 2 - g_{\phi i} \Bar{\phi}_0~,
\end{equation}

where, $B_i$, $Q_i$ and $L_i$ are baryon, charge and lepton numbers of the $i^{th}$ baryonic/leptonic species. 
$s_{\gamma}=4 \pi^2 T^3 / 45$ is the entropy density contribution from the radiation field. Entropy per baryon is then given by, $S/A=s/n_B$. Finally, we can express the energy density and pressure in the following equations,

 \begin{align}\label{eq:EoS} 
    \epsilon &= \frac{1}{2} m_{\sigma}^2 \Bar{\sigma}_0^2 + \frac{1}{2} m_{\omega}^2 \Bar{\omega}_0^2 + \frac{1}{2} m_{\rho}^2 \Bar{\rho}_{03}^2 + \frac{1}{2} m_{\phi}^2 \Bar{\phi}_0^2  \nonumber \\
    &+ U_{\sigma} + 3\Lambda_{\omega} (g_{\rho N} g_{\omega N} \Bar{\rho}_{03}.\Bar{\omega}_{0})^2 \nonumber \\
    &+ \sum_{i=B,L} \frac{2 J_i+1}{2\pi^2} \int dk k^2 E_{i}(k) \left [f_{FD}\left(\frac{E_i(k) - \mu_i^*}{T}\right)\right. \nonumber \\
    &+\left. f_{FD}\left(\frac{E_i(k) + \mu_i^*}{T}\right) \right] + \epsilon_{\gamma}~, \\
    p &= \sum_{i = B, L} \mu_i n_i + Ts - \epsilon~.
\end{align}

 We treat leptons as non-interacting Fermi gas in weak chemical equilibrium with the thermal bath and although electrons can interact with other charged species via electromagnetic interactions, we ignore these interactions as a first order approximation. The radiation represented by $A_\mu$ vector field is also treated as a non-interacting boson gas and the energy density contribution from this field is, $\epsilon_\gamma=\pi^2 T^4/15$. 

\subsection{Thermodynamic Conditions}
The various thermodynamic systems of nuclear and hypernuclear matter are described below-
\begin{enumerate}
    \item $\beta$ equilibrated/stable matter and $\nu$-free: These systems can be described by two thermodynamic variables: temperature ($T$) and density ($n_B$). Generally observed in cold NSs. The chemical potentials satisfy $\mu_e + \mu_p - \mu_n = 0$. Neutrinos are absent.
    \item Out-of-$\beta$-equilibrium and $\nu$-free: We can denote them by a set of three thermodynamic variables: ($T,~n_B,~Y_Q=Y_e$). This format is used to describe general purpose EOSs in CompOSE database~\citep{CompOSE}. The chemical potentials satisfy $\mu_e + \mu_p - \mu_n = \mu_L \neq 0$. Despite a non-zero lepton chemical potential the neutrino contribution is ignored.
    \item $\nu$-trapped matter: Described by a set of three thermodynamic variables: ($T,~n_B,~Y_L$), where $Y_L = Y_e+Y_{\nu_e}$, is the total fraction of leptons. These thermodynamic conditions are important in case of hot and dense environments such as PNS and BNS mergers~\citep{Pons_Reddy_1999apj,Raduta_Oertel_2020mnras}. The chemical potentials satisfy $\mu_e + \mu_p - \mu_n = \mu_L \neq 0$ and $\mu_{\nu_e} = \mu_L$. Neutrinos are generally trapped for $T\gtrsim5$ MeV~\citep{Alford_2018,Alford_2021}.
\end{enumerate}

In PNS early evolution the nuclear matter inside the star is temporarily opaque to neutrinos and the star goes through different stages in which the core evolves from a low temperature (low entropy state) and gets heated up (high entropy) with a decrease in lepton fraction due to neutrino emission~\citep{Prakash_Bombaci_1997phys-rep,Pons_Reddy_1999apj,Thapa_Beznogov_2023prd}. We will denote a particular hot NS configuration by the following notation: $N(S/A,Y_L)$ (for N-matter) or $NY(S/A,Y_L)$ for (NY-matter), where $S/A$ is the entropy per baryon of the configuration and $Y_L$ is the total lepton fraction. To study the effects of nuclear and hypernuclear properties on hot NS configurations, we consider the two thermal configurations $NY(1, 0.4)$ and $NY(2,0.2)$ in $\nu$-trapped regime. The former state corresponds to a hot lepton rich configuration while the latter corresponds to a hotter core with deleptonized configuration.
\\

\subsection{Thermal contribution and $\Gamma$-law} \label{subsec:gamma_law}

The $\Gamma$-law is a simplified model for hot nuclear EOS that connects it to the corresponding cold EOS. In this prescription, it is assumed that the gas is ideal and that the effects of degeneracy are negligible. We denote thermal contributions to energy density and pressure by $\epsilon_{th}$ and $p_{th}$ respectively.
\begin{align}\label{eq:thermal_contri}
    \epsilon_{th}(T,n_B,Y_Q) &= \epsilon(T,n_B,Y_Q) - \epsilon(T=0,n_B,Y_Q)~,\\
    p_{th}(T,n_B,Y_Q) &= p(T,n_B,Y_Q) - p(T=0,n_B,Y_Q)~
\end{align}
From these definitions  we can define the so-called $\Gamma$-law as,
\begin{equation}
    \Gamma_{th}(T,n_B,Y_Q) = 1 + \frac{p_{th}(T,n_B,Y_Q)}{\epsilon_{th}(T,n_B,Y_Q)}
\end{equation}

Different values of $\Gamma_{th}$ are assumed for different scenarios ($1.3<\Gamma_{th}<1.75$).
This parametrized form of hot EOS has been widely applied in different hydrodynamical simulations in astrophysics~\citep{Oechslin_Janka_2007prl,Bauswein_Janka_2012prl,Takami_Rezzolla_2015prd,Bernuzzi_Dietrich_2015prl,Blacker_Kochankovski_2024prd}. However, for EOSs involving non-nucleonic d. o. f., this prescription was shown to be not appropriate~\citep{Raduta_2022epja,Blacker_Kochankovski_2024prd}. Also, $\Gamma_{th}$ is a non-trivial varying function of density and we show this variation for our model in N- and NY-matter in Sec.~\ref{subsec:nu_free}. 

\subsection{Parameters of the model}
\label{subsec:constraints}

In this section, we describe the Bayesian scheme developed in our previous studies~\citep{Ghosh_Chatterjee_2022epja,Ghosh_Pradhan_2022_fspas,Shirke_Ghosh_2023apj,Barman_Pradhan_2025prd}
that is also employed in this work. For the Bayesian analysis, the uncertainty ranges in nuclear and hypernuclear parameters is first considered as a uniform prior. The parameter space of allowed uncertainties in nuclear and hypernuclear parameters for cold and $\beta$ stable NSs was systematically investigated in~\cite{Ghosh_Pradhan_2022_fspas}, which we also consider in this work. This parameter range considered is provided in the bottom row of Table~\ref{tab:prior} marked ``Varied". Among the hyperon potentials, the $\Lambda$-potential is relatively well-known and its value was fixed to -30 MeV~\citep{Millener_Dover_1988prc,Schaffner_Greiner_1992prc}. On the other hand, hyperon-nucleon interaction potentials for $\Sigma$- and $\Xi$-hyperons are not well constrained and we vary them within the same uncertainty range as in ~\cite{Ghosh_Pradhan_2022_fspas}. \\

Then the Bayesian cut-off filter scheme is applied to eliminate sets of parameters which do not pass through required filters. Constraints are imposed using information from microscopic Chiral Effective Field ($\chi$EFT) at low densities, Astronomical Observations (Astro) at high densities and Heavy Ion Collision (HIC) data at intermediate densities, as described below: \\
\begin{enumerate}
    \item \textbf{Chiral Effective Field Theory ($\chi$EFT):} It describes low energy interactions of protons, neutrons and pions based on symmetries of Quantum Chromodynamics (QCD). Constraints from $\chi$EFT are based on microscopic calculations from the works of \cite{Drischler_Carbone_2016prc,Drischler_Hebeler_2019prl}. They take into account long-range pion exchange interactions and low momentum expansion of nuclear forces. These constraints are available at the low density range $0.07-0.20~$fm$^{-3}$ for cold Pure Neutron Matter (PNM). \\
    \item \textbf{Astrophysical Observations (Astro):} We impose astrophysical constraints in which the maximum mass supported by a cold neutron star has a lower limit of 2.01 $M_{\odot}$, compatible with the current data (PSR J0740-6620). In addition we have constraints from gravitational wave observation of binary neutron star (BNS) merger event GW170817~\citep{LSC_2017_GW170817}. This observation  puts an upper limit on effective tidal deformability ($\tilde{\Lambda}$) to be 720 \citep{Tong_Zhao_2020prc} and radius of a 1.4 M$_{\odot}$ neutron star to be 13.5 km \citep{Most_Weih_2018prl,Eemeli_Tyler_2018prl}. \\
    \item \textbf{Heavy Ion Collisions (HIC):} HIC give insights into strong nuclear force of QCD and phase transition between different forms of matter. The constraints we use in this study have been derived from heavy-ion experiments at GSI in Germany, namely KaoS \citep{Hartnack_Oeschler_2006prl}, FOPI \citep{Fèvre_Leifels_2016npa} and ASY-EOS \citep{Russotto_2016prc}. These constraints give information about the nculear matter at intermediate density in the range 1 - 3~$n_{sat}$. The KaoS experiment can be used to put upper limits on nucleon potential ($U_N$) for ANM in the density range 2 - 3~$n_{sat}$. The FOPI experiment can be used to constrain binding energy per nucleon ($E/A$) of Symmetric Nuclear Matter (SNM) in the density range 1.4 - 2~$n_{sat}$, while ASY-EOS results put bounds on the Symmetry Energy of ANM in the supra-saturation density range 1.1 - 2~$n_{sat}$.
\end{enumerate} 

\begin{table*}
    \centering
    \caption{
    Parameter range of nuclear and hypernuclear empirical properties used in this study. The top row is a fixed set of parameters used in the study of thermal effects only while the bottom row is used in the full Bayesian analysis.}
    \label{tab:prior}
    \begin{tabular}{lccccccccc}
        \toprule
        \toprule
         &$n_{sat}$ & $E_{sat}$ & $K_{sat}$ & $J_{sym}$ & $L_{sym}$ & $m^*/m$ & $U^N_{\Lambda}$ & $U^N_{\Sigma}$ & $U^N_{\Xi}$ \\
         &($\rm{fm^{-3}}$) & (MeV) & (MeV) & (MeV) & (MeV) &  & (MeV) & (MeV) & (MeV) \\
        \midrule
        Fixed & 0.15 & -16.0 & 240 & 32 & 60 & 0.65 & -30 & 30 & -18 \\
        Varied & 0.14 to 0.17 & -16$\pm$0.2 & 200 to 300 & 28 to 34 & 40 to 70 & 0.55 to 0.75 & -30 & 0 to 30 & -30 to 0 \\
        \bottomrule
        \bottomrule
    \end{tabular}
\end{table*} 

\begin{figure*}
    \centering
    \begin{subfigure}{.32\linewidth}
        \centering
        \includegraphics[width=\textwidth]{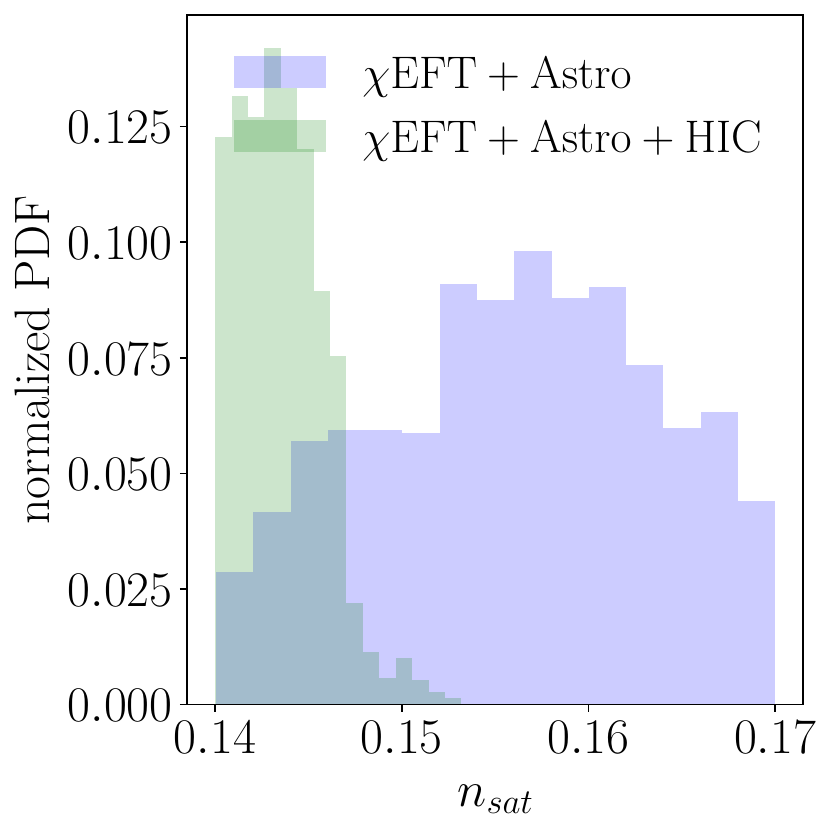}
        \caption{}
        \label{subfig:n_sat_posterior}
    \end{subfigure}
    \begin{subfigure}{.32\linewidth}
        \centering
        \includegraphics[width=\textwidth]{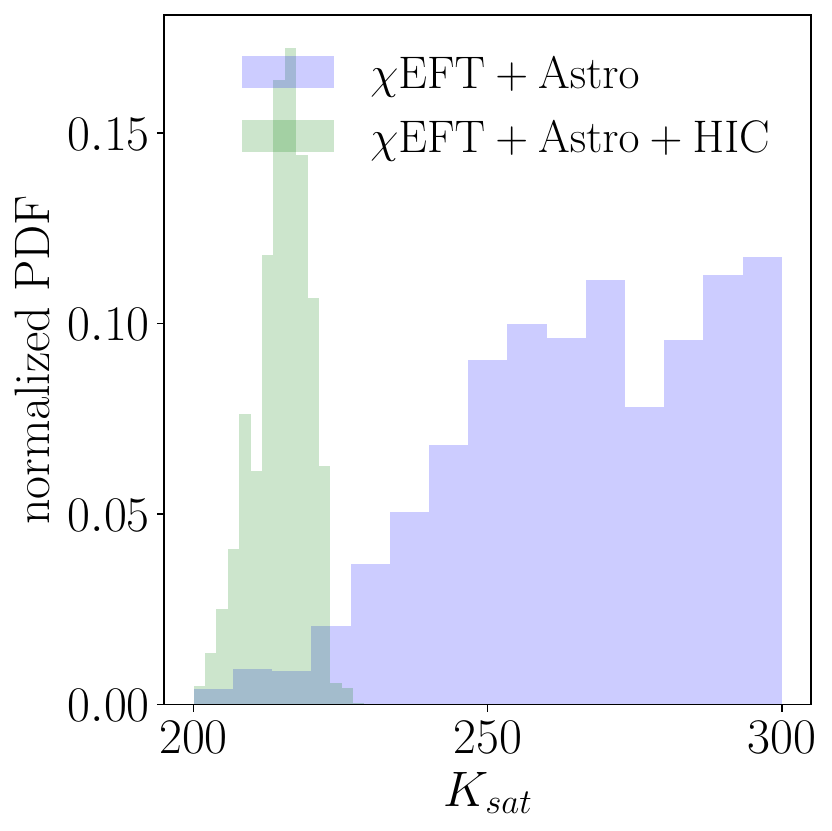}
        \caption{}
        \label{subfig:K_sat_posterior}
    \end{subfigure}
    \begin{subfigure}{.32\linewidth}
        \centering
        \includegraphics[width=\textwidth]{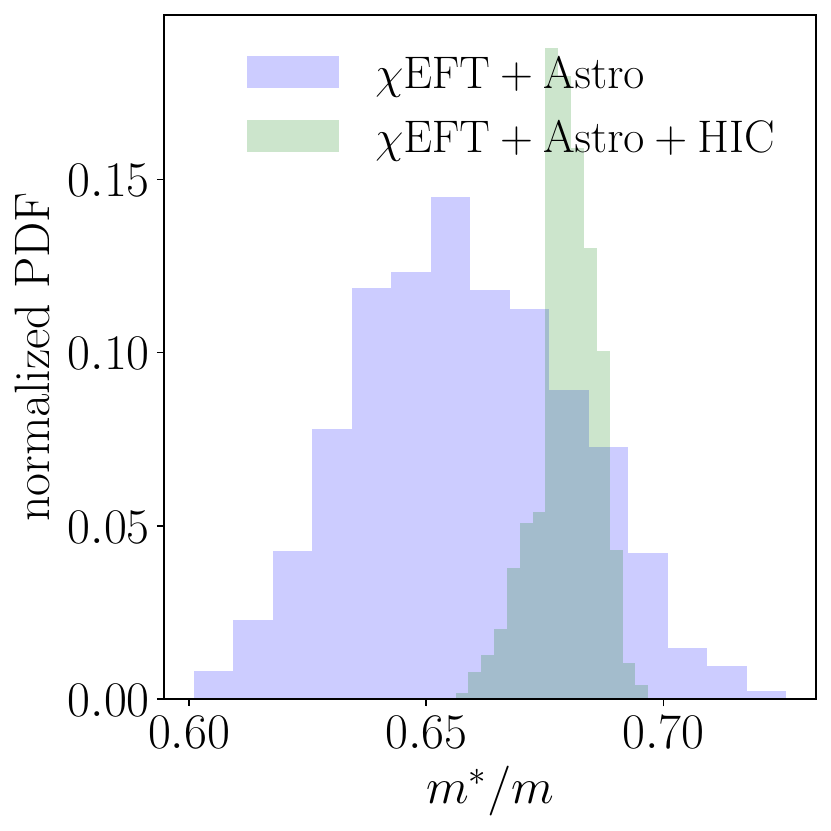}
        \caption{}
        \label{subfig:m_eff_posterior}
    \end{subfigure}
    \caption{Posteriors of isoscalar nuclear parameters with different constraints for NY-matter. (a) $n_{sat}$ posteriors, (b) $K_{sat}$ posteriors, (c) $m^*/m$ posteriors}
    \label{fig:nuclear_posterior}
\end{figure*}

After imposition of these filters, we obtain the posterior sets. In this study, we generate about 1700 posterior sets compatible with $\chi$EFT + Astro and additional $\sim$ 2300 posterior sets constrained by $\chi$EFT + Astro + HIC.
We display the posterior sets for isoscalar nuclear parameters in Fig.~\ref{fig:nuclear_posterior}. 
We find that the posteriors for saturation density, nucleon effective mass and incompressibility are not constrained by $\chi$EFT and Astro filters and span the entire prior range.
However, after imposing HIC filters along with these two, we see that $n_{sat}$ and $K_{sat}$ 
are only compatible with  low values while $m^*/m$ median value is higher compared to its median value with $\chi$EFT + Astro. 
We use these posteriors sets to construct EOSs for hot 
NS configurations with neutrino trapping and discuss how these properties affect macroscopic properties of those configurations in the subsequent sections.

\section{Results}
\label{sec:results}

\subsection{$\nu$-free case: Isothermal configuration}\label{subsec:nu_free}
\begin{figure*}
    \centering
    \begin{subfigure}{.45\linewidth}
        \centering
        \includegraphics[width=\textwidth]{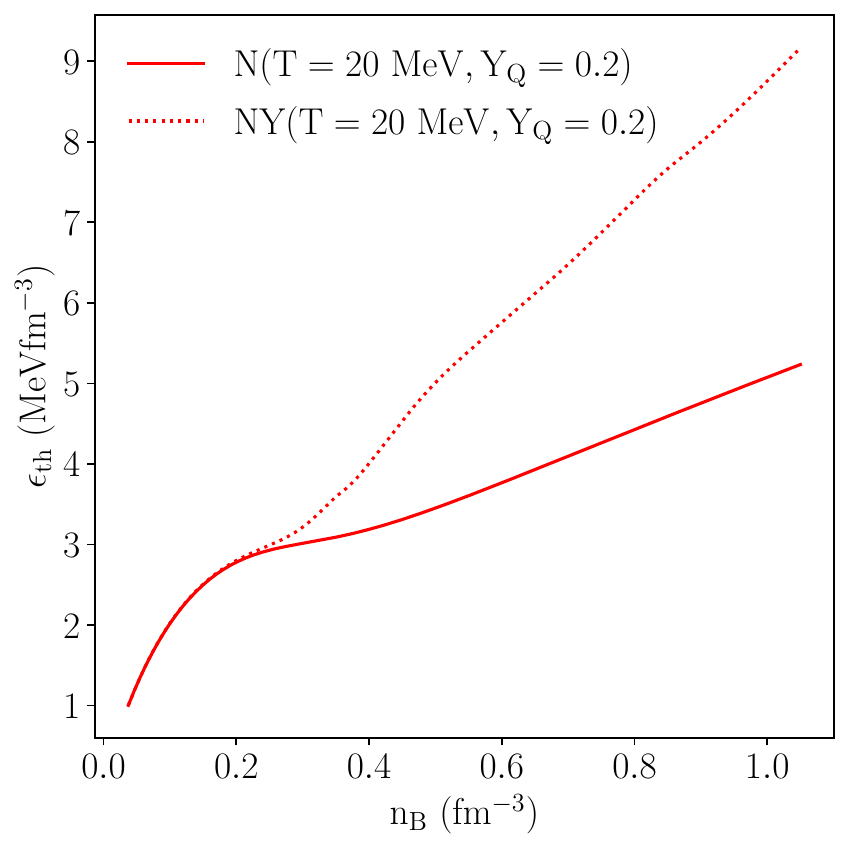}
        \caption{}
        \label{subfig:enden_th_nb_t_20_y_0.2}
    \end{subfigure}
    \begin{subfigure}{.46\linewidth}
        \centering
        \includegraphics[width=\textwidth]{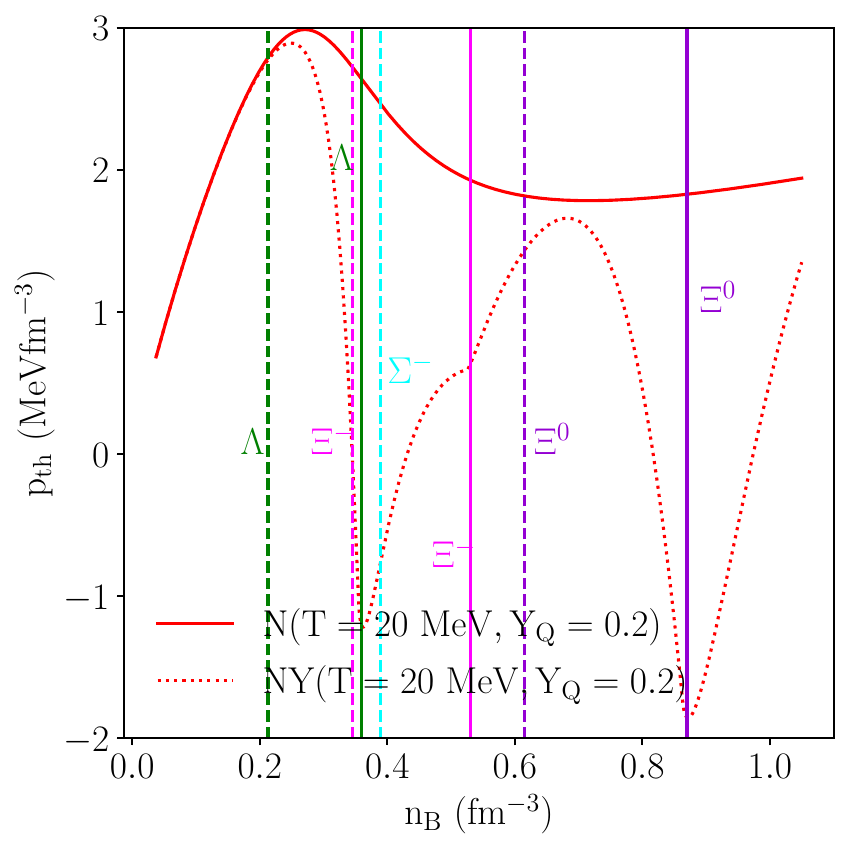}
        \caption{}
        \label{subfig:pres_th_nb_t_20_y_0.2}
    \end{subfigure}
    \caption{Thermal contributions to energy density and pressure. Nuclear and hypernuclear properties are set to fixed values mentioned in Table~\ref{tab:prior} (a) $\epsilon_{th}$ vs $n_B$, (b) $p_{th}$ vs $n_B$ [Hyperon thresholds are shown with vertical lines: dashed(solid) lines correspond to appearance in case of hot(cold) matter]} 
    \label{fig:enden_pres_th_nb_t_20_y_0.2}
\end{figure*}

\begin{figure*}
    \centering
    \begin{subfigure}{.45\linewidth}
        \centering
        \includegraphics[width=\textwidth]{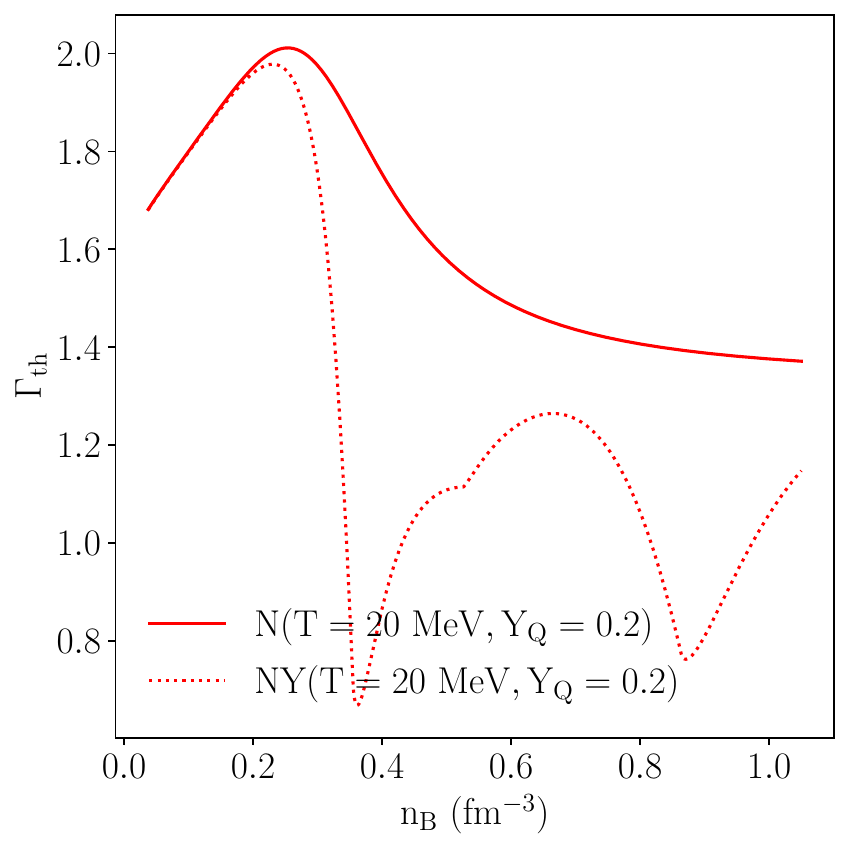}
        \caption{}
        \label{subfig:gamma_th_nb_t_20_y_0.2}
    \end{subfigure}
    \begin{subfigure}{.54\linewidth}
        \centering
        \includegraphics[width=\textwidth]{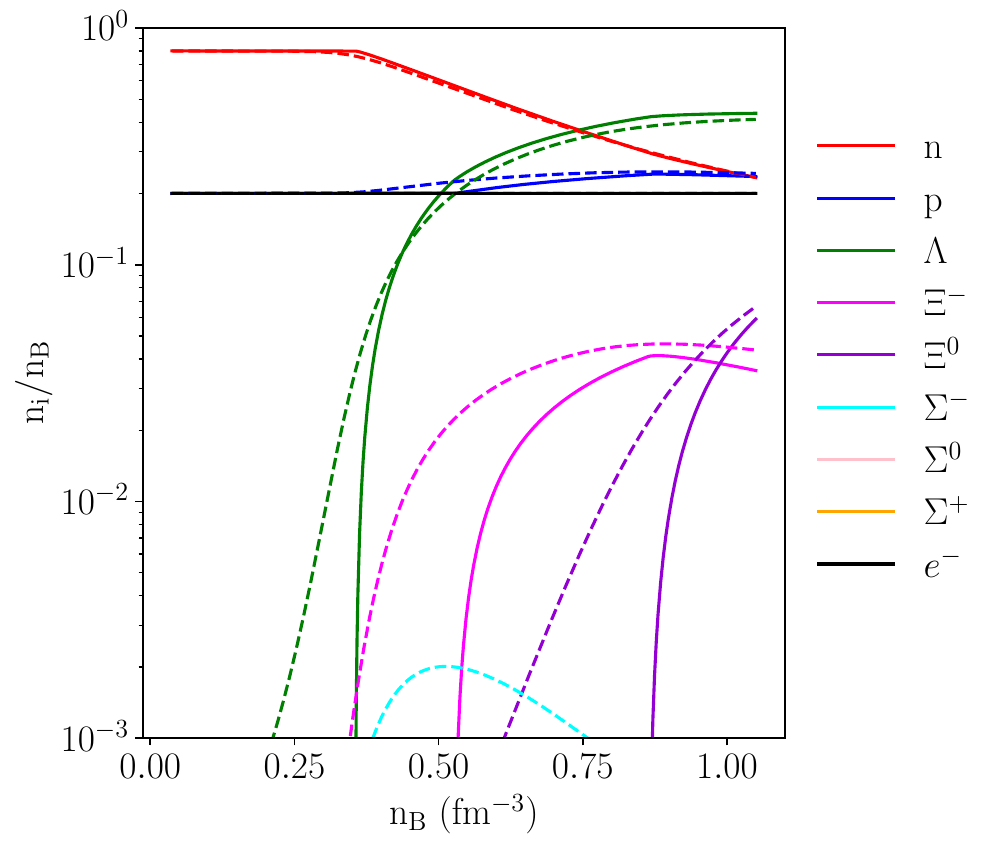}
        \caption{}
        \label{subfig:NY_frac_nb_t_0_20_y_0.2}
    \end{subfigure}
    \caption{$\Gamma$-law and particle fractions. Nuclear and hypernuclear properties are set to fixed values mentioned in Table~\ref{tab:prior} (a) $\Gamma_{th}$ vs $n_B$, (b) particle fractions vs $n_B$ [Solid (dashed) curves are for $T=0~(T=20~\rm{MeV})$, $Y_Q=0.2$ case(s)]}
    \label{fig:eos_frac_nb_t_20_y_0.2}
\end{figure*}

 We first compare hot and cold NS configurations for N- and NY-matter without neutrinos. For this, we consider two thermodynamic conditions: $T=0$ MeV and $Y_Q=0.2$; $T=20$ MeV and $Y_Q=0.2$. In order to isolate the effects of thermal effects from the uncertainties in nuclear and hypernuclear properties, we first use a fixed nuclear parametrization (top row in Table~\ref{tab:prior}) and vary only the thermal properties of the star as in~\cite{Barman_Pradhan_2025prd}.
The particle fractions as a function of baryon density are displayed in Fig.~\ref{subfig:NY_frac_nb_t_0_20_y_0.2}, from which it is seen that hyperonic species appear at even lower densities in case of finite temperature compared to zero temperature. This result is consistent with previous studies of other hyperonic finite temperature EOSs~\citep{Fortin_Oertel_2018pasa,Stone_Dexheimer_2021mnras,Raduta_Oertel_2020mnras,Sedrakian_Harutyunyan_2021universe,Tsiopelas_Sedrakian_2024}.
The thermal contribution to the energy density is monotonic, however $\epsilon_{th}$ rises more steeply with baryon density from around 0.2 $\rm{fm^{-3}}$.  From the composition plot for these two cases in Fig.~\ref{subfig:NY_frac_nb_t_0_20_y_0.2}, we observe that $\Lambda$ species appear at this same density and it explains the behavior of $\epsilon_{th}$. To understand the trend followed by $p_{th}$, we show the threshold densities of appearance of various hyperonic species in Fig.~\ref{subfig:pres_th_nb_t_20_y_0.2}. When a new species appears in the hot NY-matter system, $p_{\rm{th}}$ starts to decline with density. The dips/kinks in $p_{th}$ correspond to appearance of a hyperonic species in cold NY-system. These densities are: $\sim0.35,0.50,0.87~\rm{fm^{-3}}$. We also show $\Gamma_{th}$ as a function of baryon density in Fig.~\ref{subfig:gamma_th_nb_t_20_y_0.2}, and we can see that $\Gamma_{th}$ dips when a hyperonic species appear in the cold NY-matter at the corresponding density. As discussed in~\ref{subsec:gamma_law}, our results therefore confirm that the $\Gamma$-law description is not appropriate to describe finite temperature NY-matter.

\subsection{Isentropic Configuration: $\nu$-free vs $\nu$-trapped} \label{subsec:nu_free_vs_trapped}

\begin{figure*}
    \centering
    \begin{subfigure}{.45\linewidth}
        \centering
        \includegraphics[width=\textwidth]{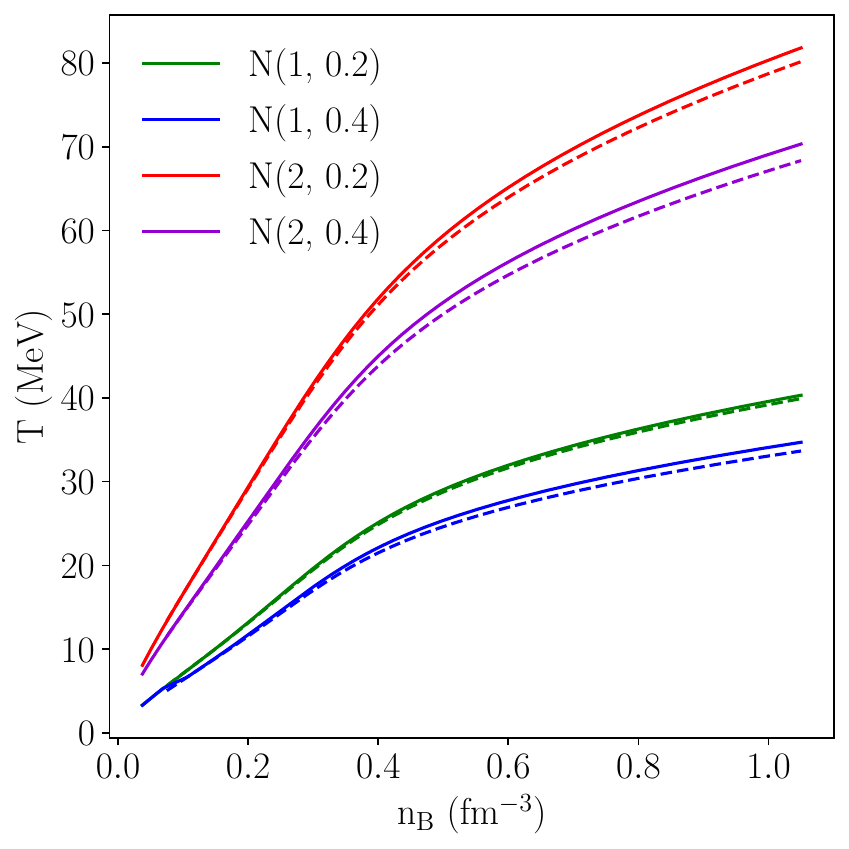}
        \caption{}
        \label{subfig:N_T_vs_nb}
    \end{subfigure}
    \begin{subfigure}{.46\linewidth}
        \centering
        \includegraphics[width=\textwidth]{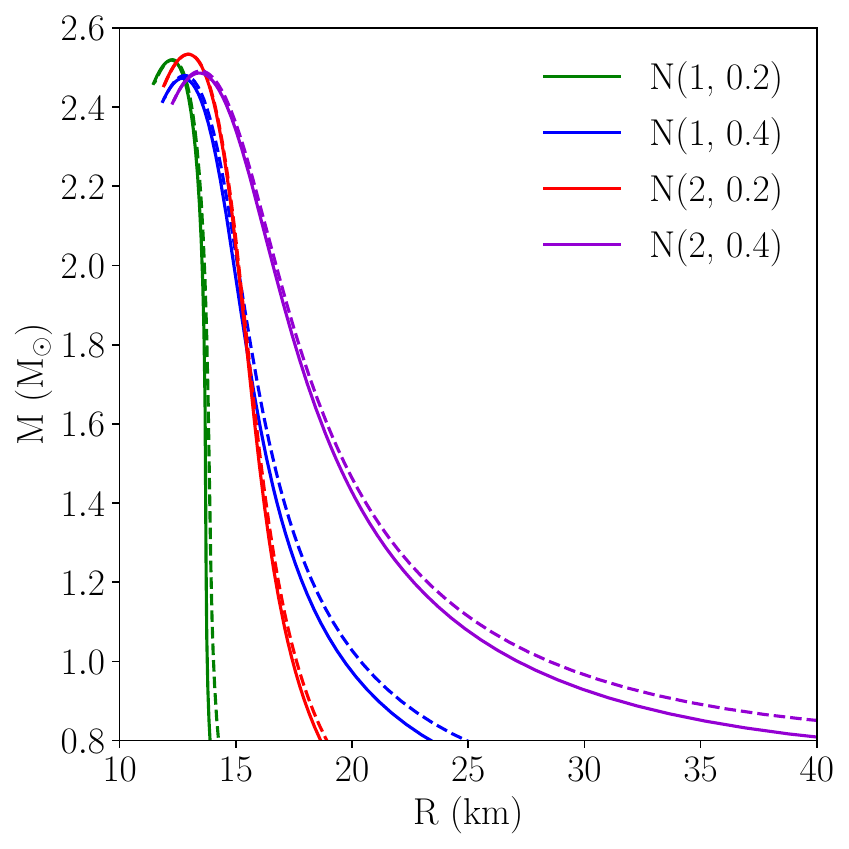}
        \caption{}
        \label{subfig:N_m_vs_r}
    \end{subfigure}
    \caption{Comparison between $\nu$-free and $\nu$-trapped cases for nucleonic (N-) matter. Solid (dashed) curves indicate $\nu$-free ($\nu$-trapped) cases. The nuclear properties are set to the fixed values mentioned in Table~\ref{tab:prior}. (a) Temperature profiles, (b) Mass-Radius relations}
    \label{fig:N_T_prof_mr}
\end{figure*}

\begin{figure*}
    \centering
    \begin{subfigure}{.45\linewidth}
        \centering
        \includegraphics[width=\textwidth]{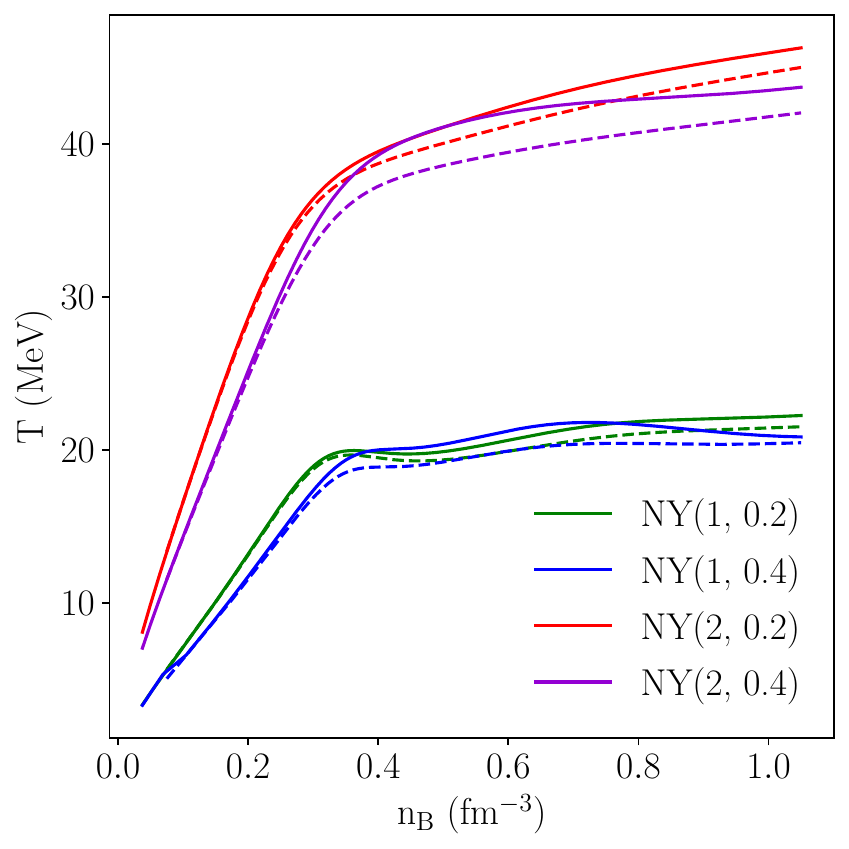}
        \caption{}
        \label{subfig:NY_T_vs_nb}
    \end{subfigure}
    \begin{subfigure}{.46\linewidth}
        \centering
        \includegraphics[width=\textwidth]{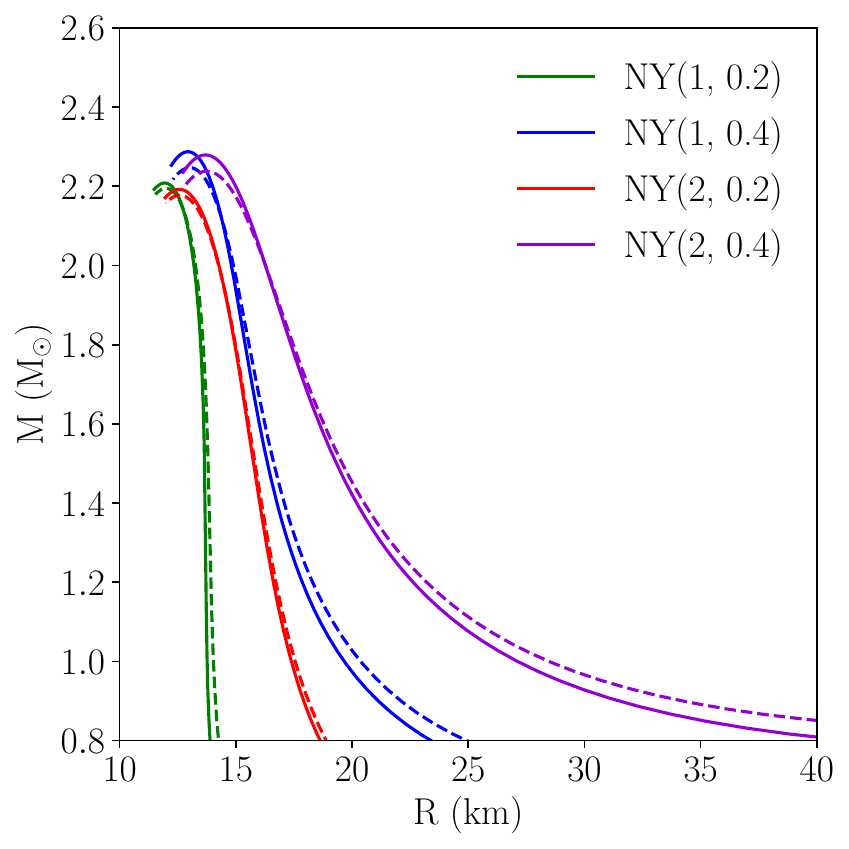}
        \caption{}
        \label{subfig:NY_m_vs_r}
    \end{subfigure}
    \caption{Comparison between $\nu$-free and $\nu$-trapped cases for hyperonic (NY-) matter. Solid (dashed) curves indicate $\nu$-free ($\nu$-trapped) cases. The nuclear and hypernuclear properties are set to the fixed values mentioned in Table~\ref{tab:prior}. (a) Temperature profiles, (b) Mass-Radius relations}
    \label{fig:NY_T_prof_mr}
\end{figure*}

As isothermal NS configurations are known to be unstable, we study isentropic configurations for neutrino-free and neutrino-trapped cases.
The lepton fraction ($Y_L$) for neutrino free configurations are same as its charge fraction ($Y_Q$). However in case of neutrino trapping, $Y_L=Y_Q+Y_{\nu_e}$. 
We consider four hot NS configurations: $(1, 0.2)$, $(1, 0.4)$, $(2, 0.2)$, $(2, 0.4)$ in both regimes for N- and NY-matter. The temperature profile inside core and mass-radius relations for N-matter are shown in Fig.~\ref{fig:N_T_prof_mr}. We see some suppression of temperature due to neutrino trapping. This can be attributed to the presence of an extra degree of freedom as opposed to npe-matter (shown in solid curves). Higher lepton fraction leads to greater temperature suppression. From the mass-radius curves in Fig.~\ref{subfig:N_m_vs_r}, we see that radii increase for intermediate and low mass configurations when neutrinos are trapped and this effect become more prominent with increasing lepton fraction. There is no significant change due to $\nu$-trapping on maximum mass supported by different thermal configurations for N-matter. For NY-matter, we can see from Fig.~\ref{subfig:NY_m_vs_r} that, temperature at high densities are further suppressed due to appearance of hyperonic species. Unlike N-matter, the maximum mass supported is lowered due to neutrino trapping and this effect increases as we increase $Y_L$. We can also observe that for N-matter, the maximum mass supported by the thermal configurations in ascending order is: $N(1,0.4) < N(2,0.4) < N(1, 0.2) < N(2, 0.2)$, while for NY-matter this trend is: $NY(2,0.2) < NY(1,0.2) < NY(2, 0.4) < NY(1, 0.4)$. Hence maximum mass decreases with lepton fraction and increases with entropy per baryon for N-matter and for NY-matter this trend is opposite.

\subsection{Composition}\label{subsec:composition}
\begin{figure*}
    \centering
    \begin{subfigure}{.44\linewidth}
        \centering
        \includegraphics[width=\textwidth]{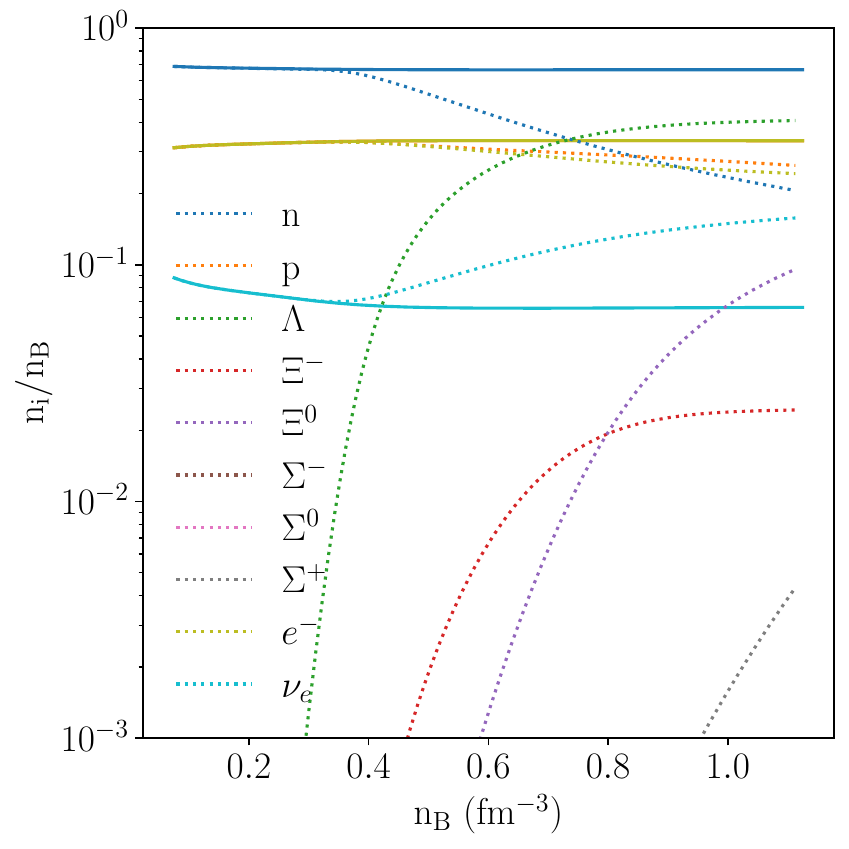}
        \caption{}
        \label{subfig:frac_1_0.4_nu}
    \end{subfigure}
    \begin{subfigure}{.44\linewidth}
        \centering
        \includegraphics[width=\textwidth]{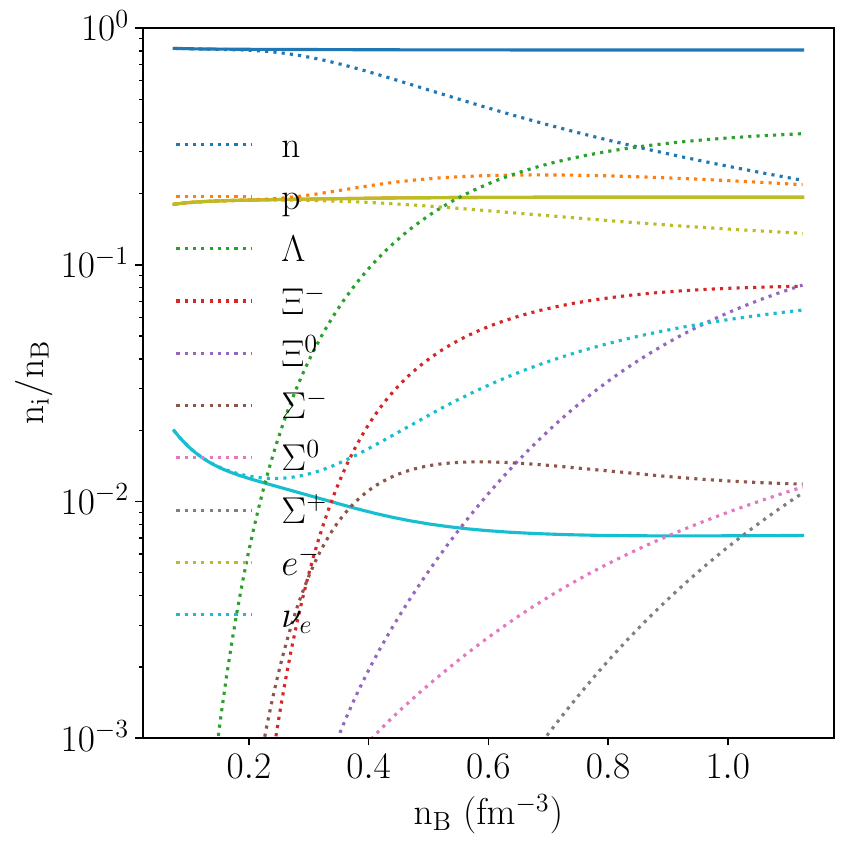}
        \caption{}
        \label{subfig:frac_2_0.2_nu}
    \end{subfigure}
    \caption{Particle fractions for N (solid)- and NY (dotted)-matter for two isentropic thermal configurations are shown. The nuclear and hypernuclear properties are set to the fixed values mentioned in Table~\ref{tab:prior}. (a) $S/A=1$, $Y_L=0.4$, (b) $S/A=2$, $Y_L=0.2$}
    \label{fig:frac_entropy_combined}
\end{figure*}

In this section we discuss the particle compositions for N- and NY-matter in $\nu$-trapped regime. We show the results for particle fractions for the two isentropic cases in Fig.~\ref{fig:frac_entropy_combined}. The different chemical species compete with each other to maintain this chemical equilibrium. When the hyperonic species start appearing, we see a decrease in fractions for both neutrons and protons and the electron fraction also goes down. As a result $\nu_e$ fraction goes up and it is further enhanced by appearance of negatively charged heavier baryons at high densities. The temperature reached at a given density is higher for $NY(2,0.2)$ case and as a result heavier baryons appear at lower densities. We can also see the thermal excitation of $\Sigma^-$ at around 0.21 $\rm{fm^{-3}}$ for this case as opposed to no appearance of this species in $N(1,0.4)$ case. The higher lepton fraction in the latter configuration also facilitates a higher neutrino fraction. The $\nu$-fraction in this case can reach $\sim15\%$ at high density as opposed to the other case ($\sim9\%$).

\subsection{Effects of constraints on hot NS configurations and $f$-mode oscillations} \label{subsec:PNS_constraints}
\begin{figure*}
    \centering
    \begin{subfigure}{.44\linewidth}
        \centering
        \includegraphics[width=\textwidth]{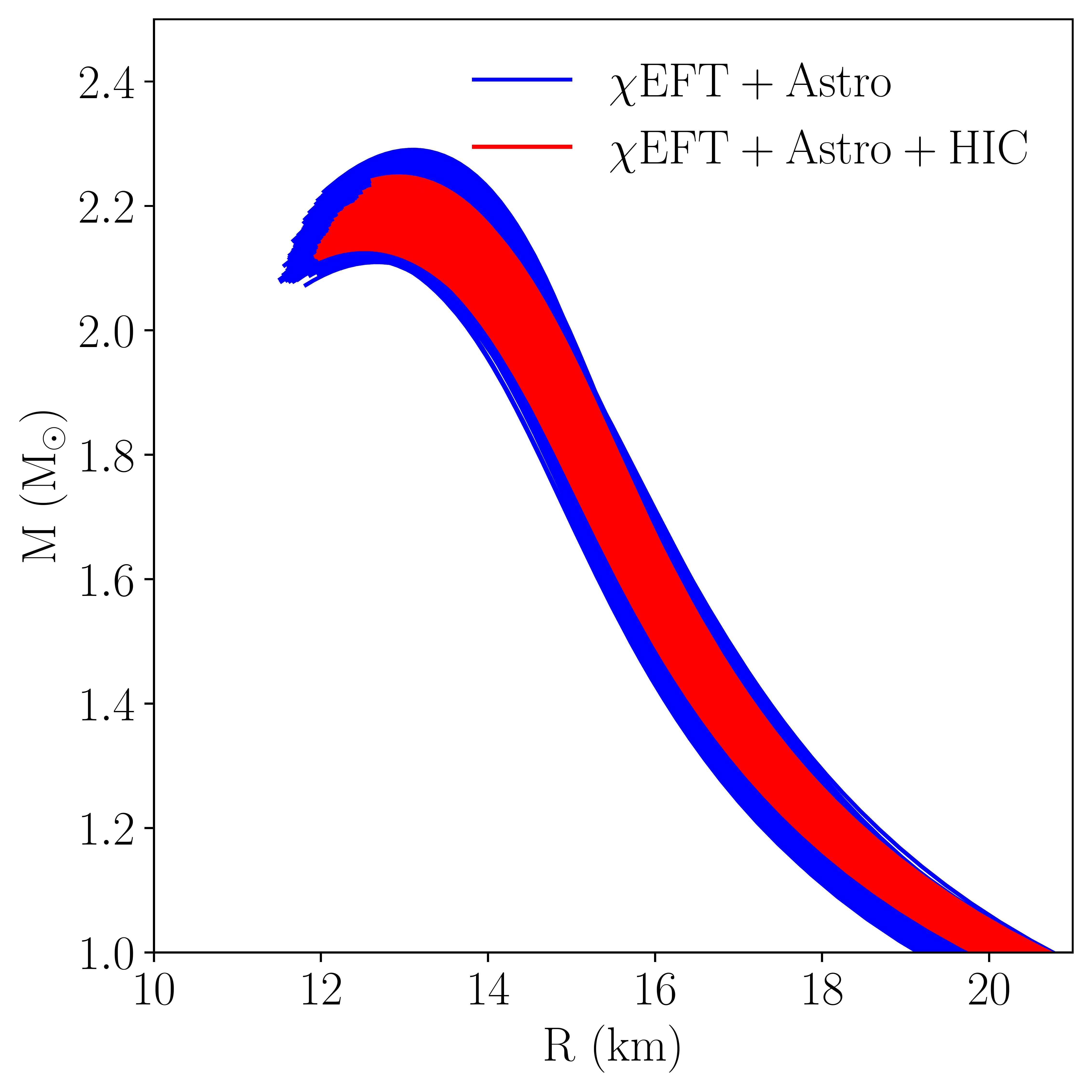}
        \caption{}
        \label{subfig:NY_1_0.4_nu_mr_combined}
    \end{subfigure}
    \begin{subfigure}{.44\linewidth}
        \centering
        \includegraphics[width=\textwidth]{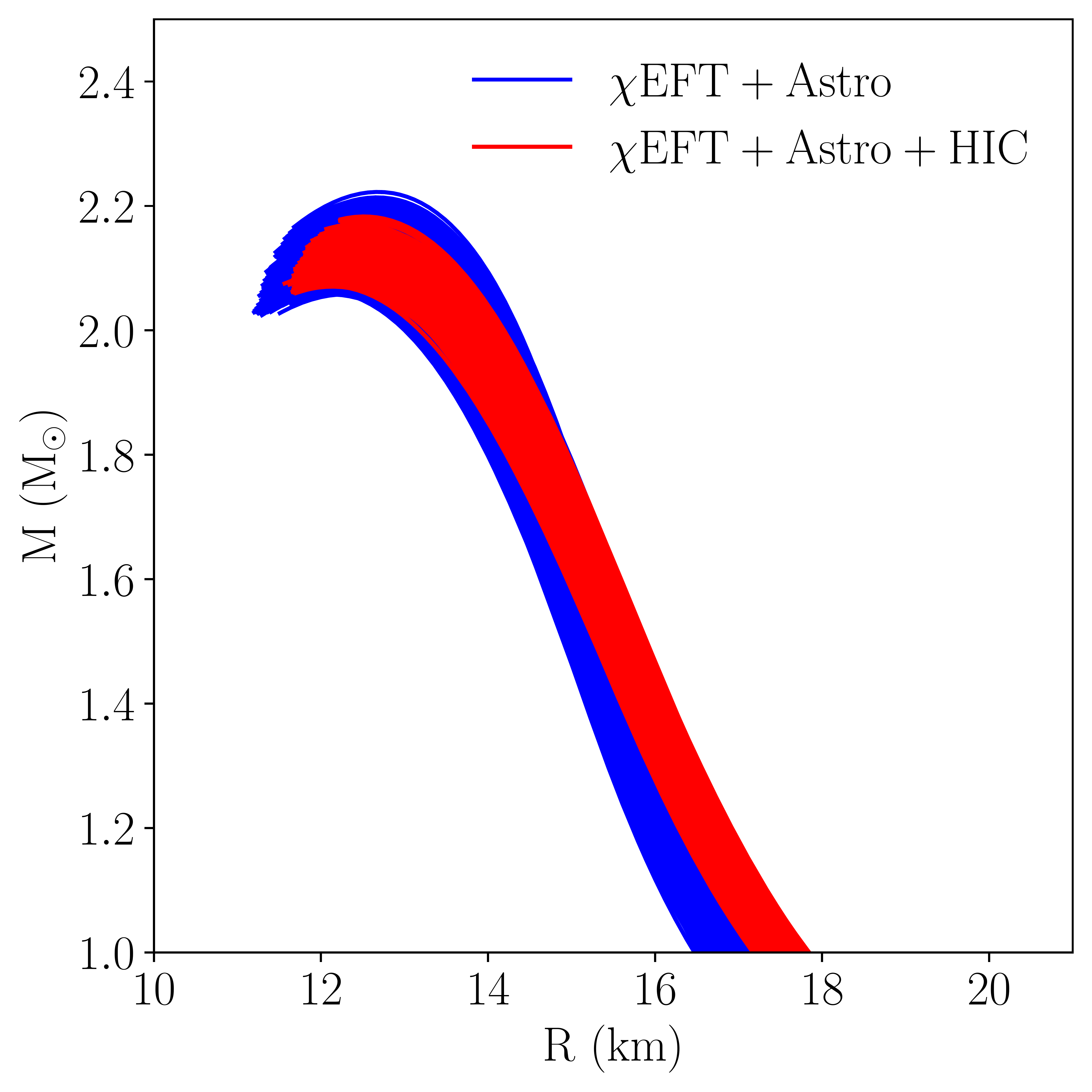}
        \caption{}
        \label{subfig:NY_2_0.2_nu_mr_combined}
    \end{subfigure}
    \caption{Mass-radius relations for two thermal configurations obtained after imposing constraints from $\chi$EFT, Astrophysical observations (Astro) and Heavy Ion Collisions (HIC). (a) $S/A=1$, $Y_L=0.4$, (b) $S/A=2$, $Y_L=0.2$}
    \label{fig:NY_nu_mr_combined}
\end{figure*}

\begin{figure*}
    \centering
    \begin{subfigure}{.44\linewidth}
        \centering
        \includegraphics[width=\textwidth]{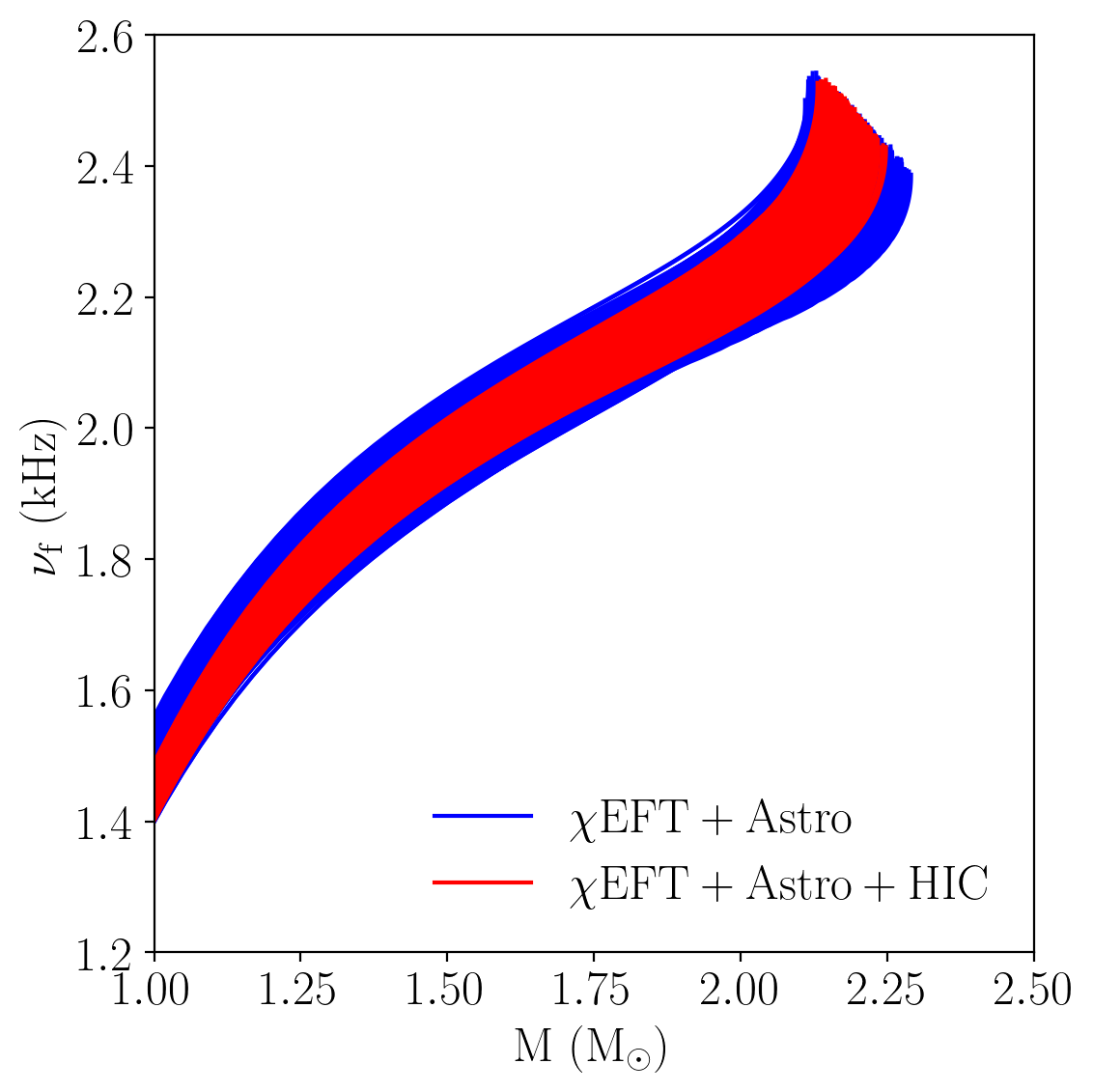}
        \caption{}
        \label{subfig:NY_1_0.4_nu_fm_combined}
    \end{subfigure}
    \begin{subfigure}{.44\linewidth}
        \centering
        \includegraphics[width=\textwidth]{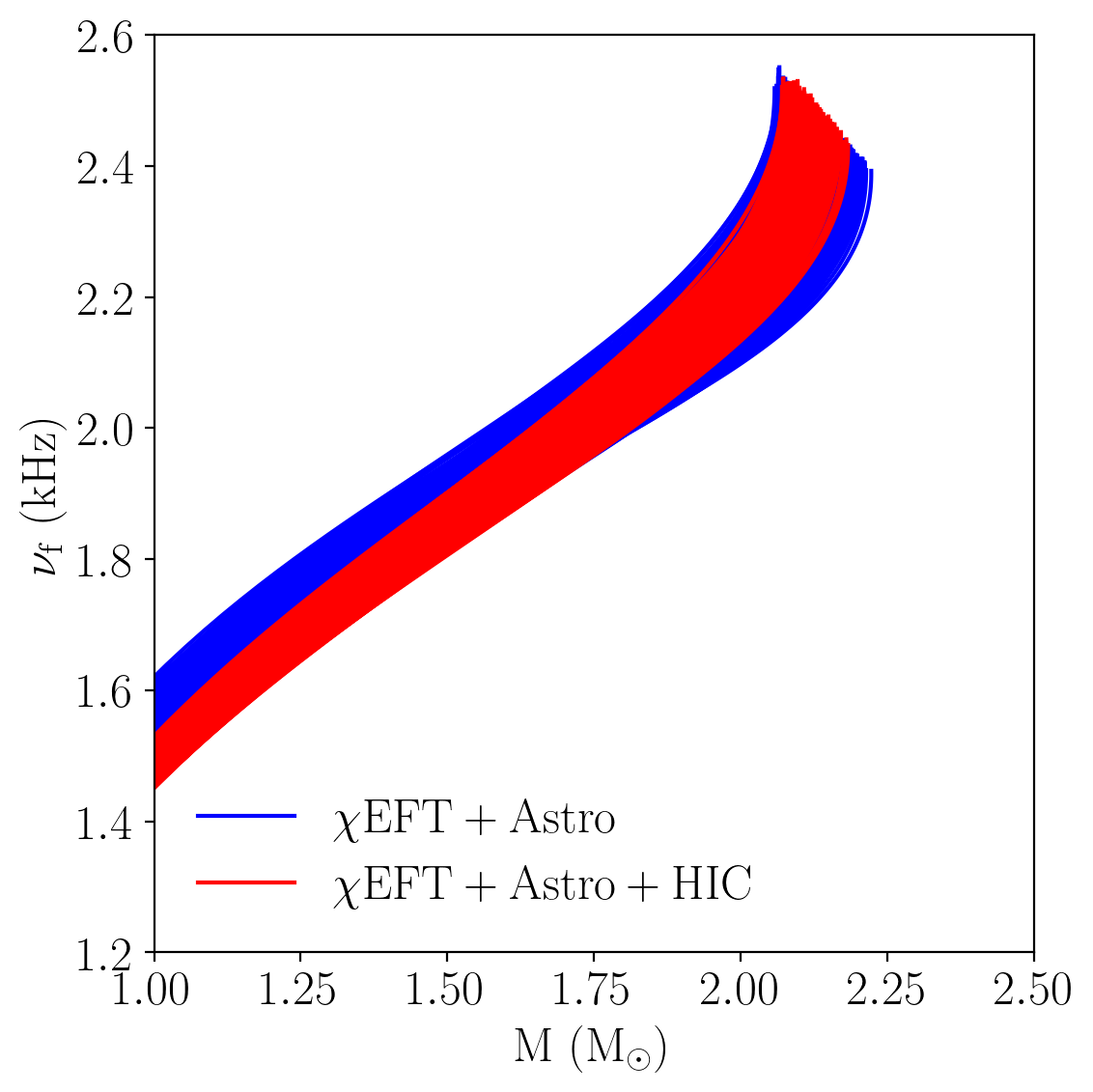}
        \caption{}
        \label{subfig:NY_2_0.2_nu_fm_combined}
    \end{subfigure}
    \caption{$f$-mode frequencies (Cowling) as a function of mass for two thermal configurations obtained after imposing constraints from $\chi$EFT, Astrophysical observations (Astro) and Heavy Ion Collisions (HIC). (a) $S/A=1$, $Y_L=0.4$, (b) $S/A=2$, $Y_L=0.2$}
    \label{fig:NY_nu_fm_combined}
\end{figure*}

We now vary all the nuclear and hypernuclear parameters to perform a full Bayesian study for the two thermal configurations: $NY(1, 0.4)$ and $NY(2,0.2)$ in $\nu$-trapped regime. The posterior mass-radius relations for these configurations are shown in Fig.~\ref{fig:NY_nu_mr_combined}. We can again see that the maximum mass supported by $NY(1,0.4)$ configurations are higher with the constraints considered. The posteriors of the lepton rich configuration support larger radii for intermediate mass configurations. Imposition of HIC constraints leads to EOSs becoming softer. We also show the posteriors of $f$-mode frequencies as a function of mass in Fig.~\ref{fig:NY_nu_fm_combined}. As in our previous work~\cite{Barman_Pradhan_2025prd}, we use Cowling approximation to calculate $f$-mode oscillation frequencies. We see that the posterior bands for the two configurations overlap for the high mass cases but they differ for intermediate mass cases. \\

We now discuss the role of nuclear and hypernuclear parameters on the macroscopic properties and $f$-mode oscillations for the two representative thermal configurations. To understand the correlations, we show joint posterior distributions of some important nuclear properties with some macroscopic properties of the aforementioned hot PNS configurations. These joint distributions are found after imposing constraints from $\chi$EFT, Astro and HIC. \\

\begin{figure*}
    \centering
    \begin{subfigure}{.44\linewidth}
        \centering
        \includegraphics[width=\textwidth]{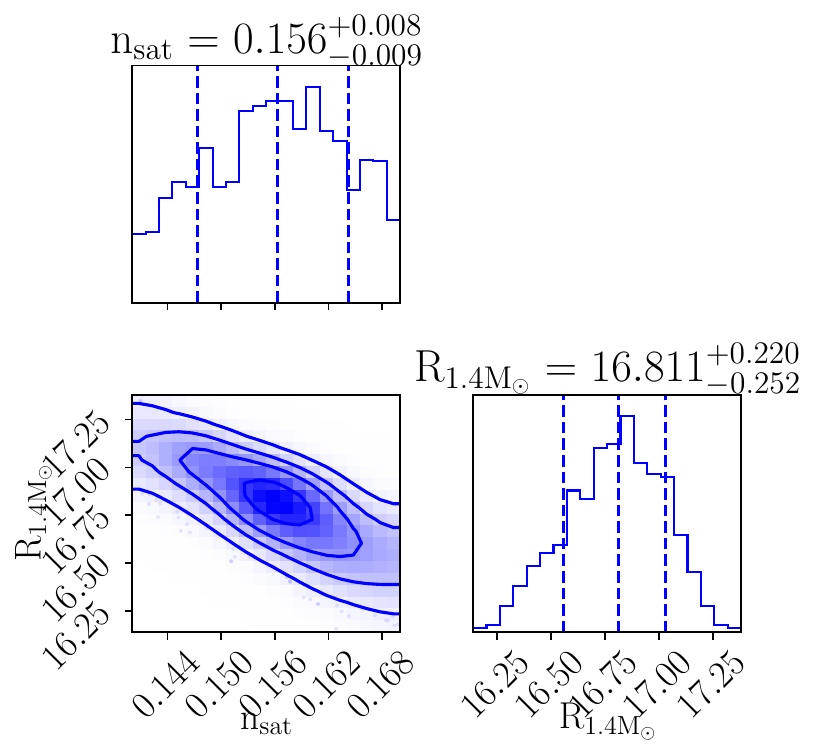}
        \caption{}
        \label{subfig:NY_1_0.4_nu_n_vs_R_1.4_corner}
    \end{subfigure}
    \begin{subfigure}{.44\linewidth}
        \centering
        \includegraphics[width=\textwidth]{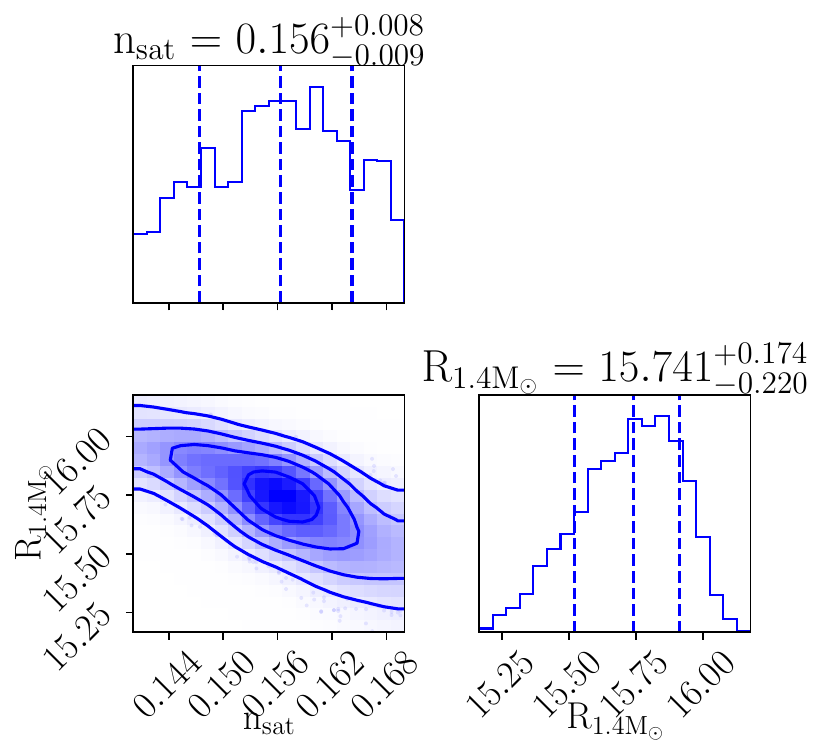}
        \caption{}
        \label{subfig:NY_2_0.2_nu_n_vs_R_1.4_corner}
    \end{subfigure}
    \caption{Joint posteriors for $n_{sat}$ (in $\rm{fm^{-3}}$) and Radius (in km) of canonical $1.4M_{\odot}$ hot NSs for two thermal configurations  with constraints from $\chi$EFT and Astro. (a) $S/A=1$, $Y_L=0.4$, (b) $S/A=2$, $Y_L=0.2$}
    \label{fig:NY_nu_n_vs_R_1.4_corner}
\end{figure*}

In Fig.~\ref{fig:NY_nu_n_vs_R_1.4_corner}, we have shown the joint posterior distributions of saturation density with radius of $1.4M_{\odot}$ canonical PNS configurations with constraints from $\chi$EFT and Astro. We observe moderate to strong anti-correlations of $n_{sat}$ with radius for both the thermal configurations. It is well known that the tidal deformability and $f$-mode frequency are correlated and anti-correlated with radius. Hence nuclear saturation density also shows moderate to strong correlations/anti-correlations with these observables (see appendix~\ref{appendix_a}). Meanwhile, $R_{1.4M_{\odot}}$ median value is on the higher side in the case $NY(1, 0.4)$ configuration.\\

The HIC constraints push saturation properties to lower values. The correlations/anti-correlations with observable properties of $1.4M_{\odot}$ canonical NS configurations still persist. Also it turns out that observables of $2M_{\odot}$ canonical PNS configurations now show moderate correlations with saturation density (see appendix~\ref{appendix_b}). In Fig.~\ref{fig:NY_nu_n_vs_f_2_corner_hic}, we show joint posteriors of $n_{sat}$ with $f$-mode frequency of $2M_{\odot}$ canonical NS configurations for both the thermal cases after imposing constraints from $\chi$EFT, Astro and HIC. The effect of these additional constraints results in $\nu_{f,2M_{\odot}}$ moderately correlating with nuclear saturation density. Also, radius and tidal deformability of $2M_{\odot}$ canonical NS configurations are moderately anti-correlated with nuclear saturation density.  This result is contrasting to hot N-matter case where it was shown that $n_{sat}$ is not at all correlated with observables of $2M_{\odot}$ canonical hot NS configurations~\citep{Barman_Pradhan_2025prd}. We also observe from Fig.~\ref{subfig:NY_1_0.4_nu_n_vs_f_2_corner_hic} and Fig.~\ref{subfig:NY_2_0.2_nu_n_vs_f_2_corner_hic} that the posterior distributions of $f$-mode frequencies are only marginally different from each other for the two thermal configurations considered here.  \\

\begin{figure*}
    \centering
    \begin{subfigure}{.45\linewidth}
        \centering
        \includegraphics[width=\textwidth]{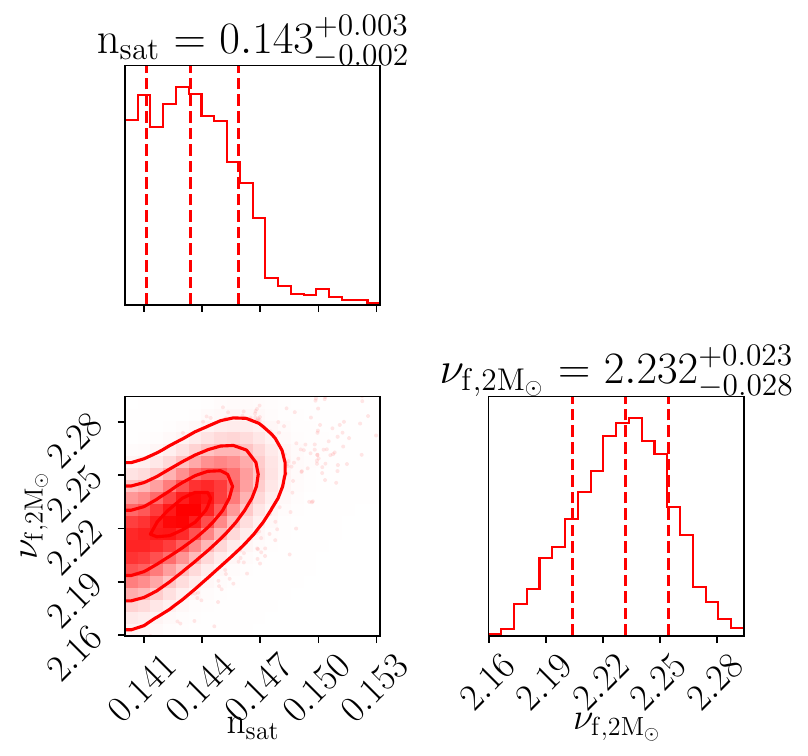}
        \caption{}
        \label{subfig:NY_1_0.4_nu_n_vs_f_2_corner_hic}
    \end{subfigure}
    \begin{subfigure}{.45\linewidth}
        \centering
        \includegraphics[width=\textwidth]{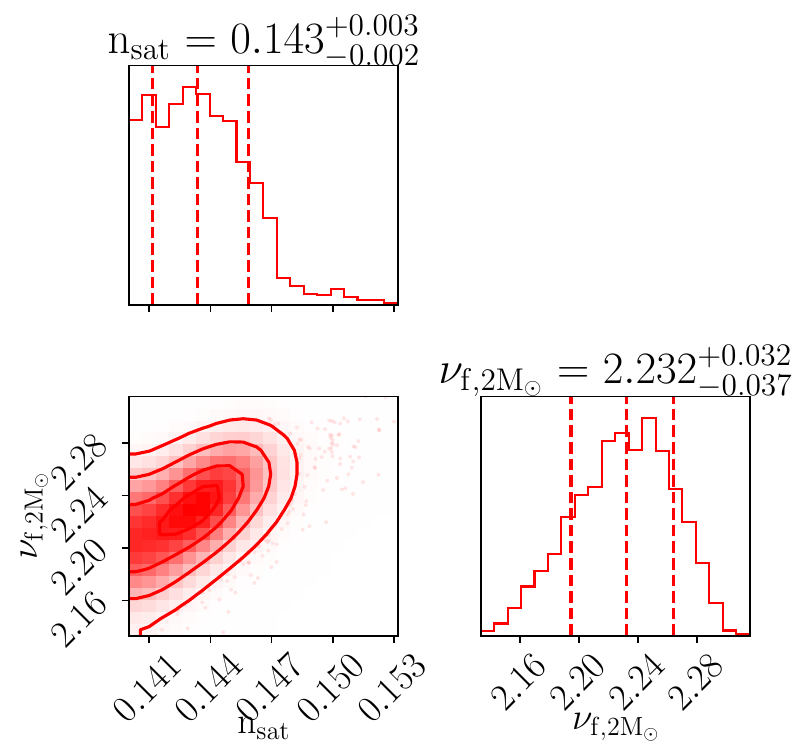}
        \caption{}
        \label{subfig:NY_2_0.2_nu_n_vs_f_2_corner_hic}
    \end{subfigure}
    \caption{Joint posteriors for $n_{sat}$ (in $\rm{fm^{-3}}$) and $f$-mode frequency (in kHz) of canonical $2M_{\odot}$ hot NSs for two thermal configurations with constraints from $\chi$EFT, Astro and HIC. (a) $S/A=1$, $Y_L=0.4$, (b) $S/A=2$, $Y_L=0.2$}
    \label{fig:NY_nu_n_vs_f_2_corner_hic}
\end{figure*}

\begin{figure*}
    \centering
    \begin{subfigure}{.45\linewidth}
        \centering
        \includegraphics[width=\textwidth]{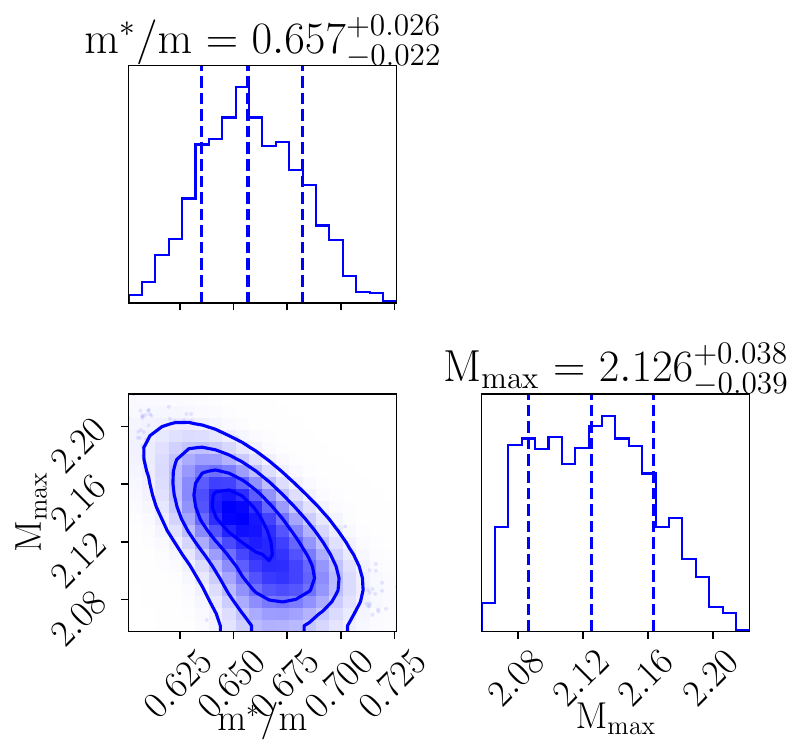}
        \caption{}
        \label{subfig:NY_2_0.2_nu_m_vs_M_max_corner}
    \end{subfigure}
    \begin{subfigure}{.45\linewidth}
        \centering
        \includegraphics[width=\textwidth]{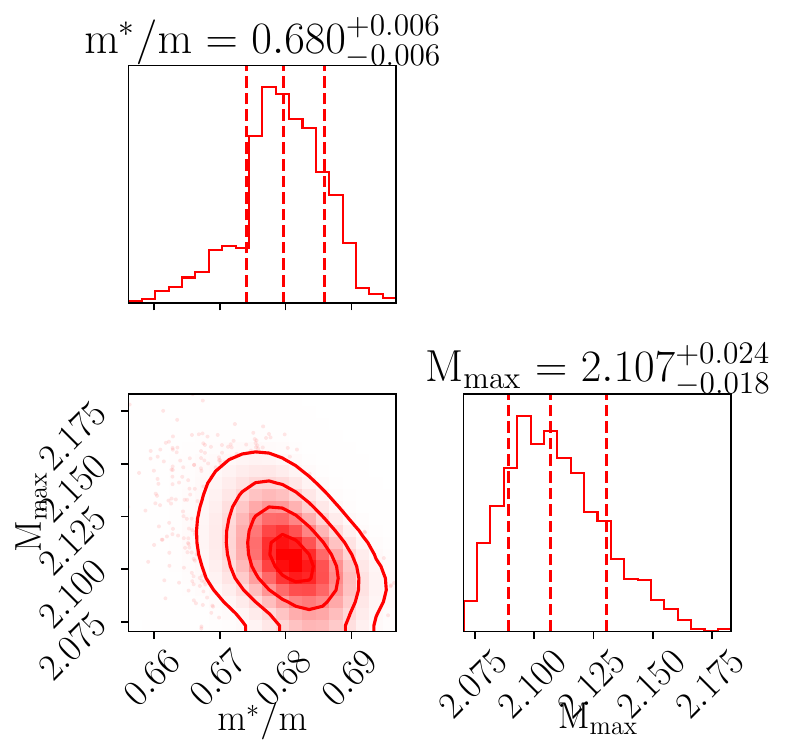}
        \caption{}
        \label{subfig:NY_2_0.2_nu_m_vs_M_max_corner_hic}
    \end{subfigure}
    \caption{Joint posteriors for effective mass and maximum mass in $M_{\odot}$ of hot NSs for $NY(2,0.2)$ configuration. (a) $\chi$EFT + Astro (correlation = - 0.66), (b) $\chi$EFT + Astro + HIC (correlation = - 0.47)}
    \label{fig:NY_nu_m_M_max_corner_all}
\end{figure*}

The next important nuclear property that we found to affect the stiffness of the EOS is the nucleon effective mass. We show the joint posteriors of this parameter with maximum mass ($M_{max}$) supported for $NY(2, 0.2)$ configuration in Fig.~\ref{fig:NY_nu_m_M_max_corner_all}. $M_{max}$ is found to be moderately anti-correlated with $m^*/m$ but this effect decreases upon imposing HIC constraints. We have indicated in the figure caption the Pearson correlation coefficients for the respective filters.  Apart from this nucleon effective mass is found to be not correlating with any other observable properties for both thermal configurations. These results are in contrast to our previous findings for N-matter in which $M_{max}$ and $2M_{\odot}$ canonical hot NS observables showed strong correlation/anti-correlation with $m^*/m$~\citep{Barman_Pradhan_2025prd}. \\

These results are in agreement with our previous study in~\cite{Ghosh_Pradhan_2022_fspas}, where the same constraints were imposed on cold and $\beta$-equilibrated NY-matter NSs described within the same EOS formalism. It was observed that in presence of hyperons, the correlation of effective nucleon mass $m^*/m$ with NS observables was weakened, while saturation density $n_{sat}$ was found to show moderate correlation with $R_{1.4M_{\odot}}$ and strong correlation with $m^*/m$. 
In contrast with~\cite{Ghosh_Pradhan_2022_fspas} where $R_{1.4M_{\odot}}$ was found to be moderately correlated with $L_{sym}$, we do not see any such correlation for hot NY-matter NS configurations. A recent study by~\cite{Tsiopelas_Sedrakian_2024} explored the effect of $L_{sym}$ on hot nuclear and hypernuclear EOSs and found that its effect diminishes with increasing temperature. The other nuclear parameters along with the hyperon potentials do not affect NS properties (see appendix~\ref{appendix_a} and~\ref{appendix_b}).

\section{Universal Relations}
Cold and $\beta$-stable neutron stars are known to exhibit certain EOS independent relations among various macroscopic properties such as moment of inertia, compactness, tidal deformability, $f$-mode frequency etc.~\mbox{\cite{Yagi_Yunes_2013prd}} showed that the moment of inertia ($I$), the dimensionless tidal deformability ($\Lambda$) and the spin-induced quadruple moment ($Q$)
are related to each other in a manner independent of the underlying nuclear EOS. This is known as $\Bar{I}-Love-\Bar{Q}$ relation and~\mbox{\cite{Maselli_Cardoso_2013prd}} in their work gave the fit relation for $\Bar{I}-\Lambda$. In addition, the $f$-mode frequency is found to correlate with other macroscopic properties such as mass, radius, average mass density, compactness and dimensionless tidal deformability~\protect\citep{Andersson_Kokkotas_1998mnras,Benhar_Ferrari_2004prd,Doneva_Gaertig_2013prd,Pradhan_Chatterjee_2022prc}.~\mbox{\cite{Maselli_Cardoso_2013prd}} also showed that compactness and dimensionless tidal deformability correlate with each other independently of EOS ($C-Love$ relation). In practice, universal relations help reduce degeneracies between astrophysical measurements and nuclear physics uncertainties, enabling tests of general relativity in the strong-field regime and offering new insights into the behavior of matter at supranuclear densities. The effect of finite temperature on these Universal Relations has been studied in some previous works~\protect\citep{Raduta_Oertel_2020mnras,Laskos-Patkos_Koliogiannis_2022_universe,Laskos-Patkos_Pavlos_2023hnpsanp,Largani_Fischer_2022mnras,Khadkikar_Raduta_2021prc}. In our previous work~\mbox{\citep{Barman_Pradhan_2025prd}}, we explored $C-Love$ relations for different thermal configurations taking into account the EOS uncertainties for $\nu$-free N-matter. In this work, we use the large set of posterior EOSs obtained for hot hyperonic neutrino trapped matter to explore two of the well known universal relations: the scaling of $f$-mode frequency with square root of average mass density and the $C-\Lambda$ relation.

\subsection{Relation between $f$-mode frequency and average mass density}

\begin{figure}
    \centering
    \includegraphics[width=\linewidth]{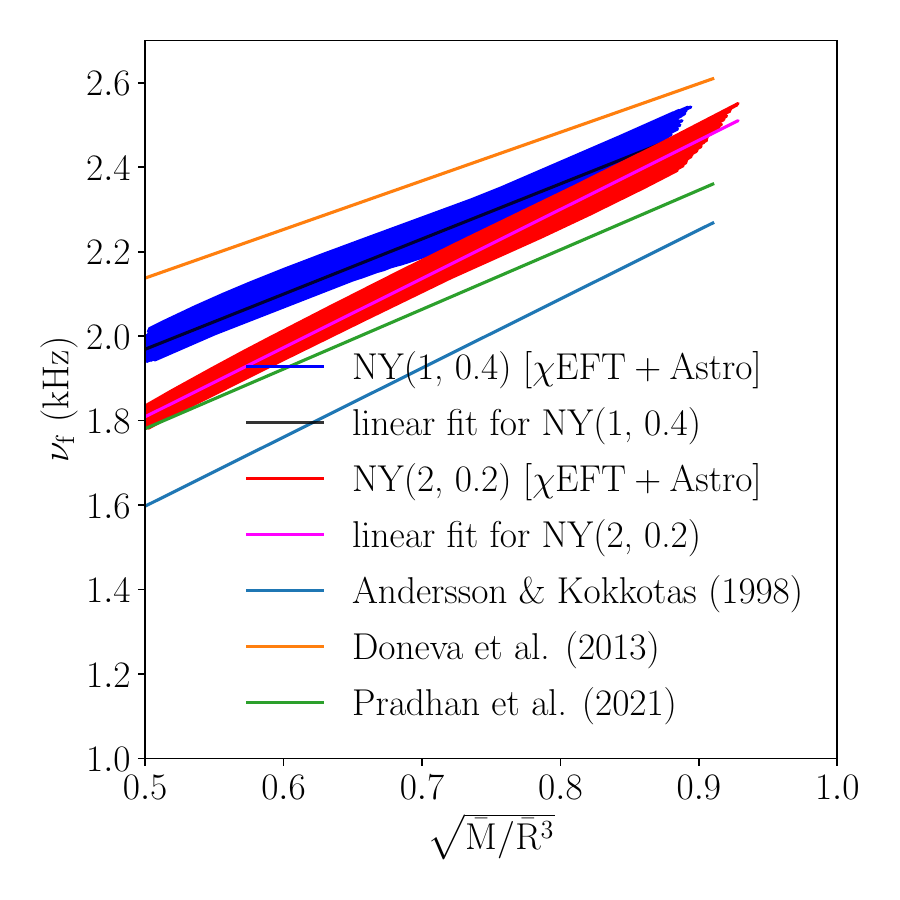}
    \caption{Relation between $f$-mode frequency and square root of average mass density for the thermal configurations: $NY(1,0.4)$ (blue) and $NY(2,0.2)$ (red). Linear fit relations found for these configurations are shown in black and magenta lines respectively. Also shown are fit relations from~\protect\cite{Andersson_Kokkotas_1998mnras,Doneva_Gaertig_2013prd,Pradhan_Chatterjee_2021prc}}
    \label{fig:f_vs_avg_den}
\end{figure}

\cite{Andersson_Kokkotas_1998mnras} in their work found that $f$-mode frequency of a cold and $\beta$ stable NS is positively correlated to the square root of the average mass density of the corresponding NS. They proposed an empirical linear relation between these two quantities as following,

\begin{equation} \label{eq:f_vs_avg_den}
    \nu_f = a + b \sqrt{\frac{\bar{M}}{\bar{R}^3}},
\end{equation} \\
where, $\bar{M}=\frac{M}{1.4M_{\odot}}$ and $\bar{R}=\frac{R}{10~\rm{km}}$ are normalized mass and radius respectively and $\nu_f$ is in kHz. Subsequent studies with more number of cold EOSs by~\cite{Doneva_Gaertig_2013prd} and~\cite{Pradhan_Chatterjee_2021prc} refined this fit relation.  We study the same relation here for the two neutrino trapped hot hyperonic thermal configurations: $NY(1,0.4)$ and $NY(2,0.2)$ constrained by $\chi$EFT and Astro. We show the plot of these relations in Fig.~\ref{fig:f_vs_avg_den}. From this figure, we can see that the $f$-mode frequency corresponding to a particular scaled average mass density is higher for $NY(1,0.4)$ configuration than that of $NY(2,0.2)$ configuration. The cold matter fits of~\cite{Andersson_Kokkotas_1998mnras} and~\cite{Pradhan_Chatterjee_2021prc} underestimate these mode frequencies for both the thermal cases while fit by~\cite{Doneva_Gaertig_2013prd} overestimates them. We provide our modified fit relations for these thermal configurations in Table~\ref{tab:f_vs_avg_den_fit}.

\begin{table*}
    \centering
    \begin{tabular}{lcc}
        \toprule
        \toprule
         &$a$ & $b$ \\
        \midrule
        \cite{Andersson_Kokkotas_1998mnras} & 0.78 & 1.635 \\
        \cite{Doneva_Gaertig_2013prd} & 1.562 & 1.151 \\
        \cite{Pradhan_Chatterjee_2021prc} & 1.075 & 1.412 \\
        $NY(1,0.4)$ [This work] & 1.316 & 1.306 \\
        $NY(2,0.2)$ [This work] & 0.991 & 1.637 \\
        \bottomrule
        \bottomrule
    \end{tabular}
    \caption{Fit relations for $f-\sqrt{\bar{M}/\bar{R}^3}$ relations found in this work for the two hot NS configurations. These relations are obtained after taking into account the uncertainty from $\chi$EFT and Astro. Also shown are the fit relations for cold NSs from~\protect\cite{Andersson_Kokkotas_1998mnras,Doneva_Gaertig_2013prd,Pradhan_Chatterjee_2021prc}.}
    \label{tab:f_vs_avg_den_fit}
\end{table*}

\subsection{$C-Love$ relations}

The universal relation between compactness and dimensionless tidal deformability in case of cold and $\beta$ equilibrated NSs are important for estimation of the size of the system from tidal properties. However during the early post-bounce evolution of PNSs, the radii and tidal deformabilities are both larger compared to its cold counterparts and using the cold universal relation on such systems may lead to incorrect estimate for the size of the systems. Also in the context of BNS post-merger remnants, the modification of $C-Love$ relation has implication on future gravitational wave detectors that probe tidal effects in hot NS remnants.
In this section we study the sensitivity of nuclear and hypernuclear parameters on the $C-Love$ relations for the aforementioned two NS configurations with NY-matter and neutrino trapping. Constraints on these hot EOSs are $\chi$EFT + Astro. This relation can be described as a second order fit of compactness as a function of $\Lambda$ given below~\citep{Maselli_Cardoso_2013prd,Barman_Pradhan_2025prd,Raduta_Oertel_2020mnras},

\begin{equation} \label{eq:C-Lambda}
    C = \sum^2_{i=0} a_i [\rm{ln(\Lambda)}]^i~.
\end{equation}

\begin{figure*}
    \centering
    \begin{subfigure}{.45\linewidth}
        \centering
        \includegraphics[width=\textwidth]{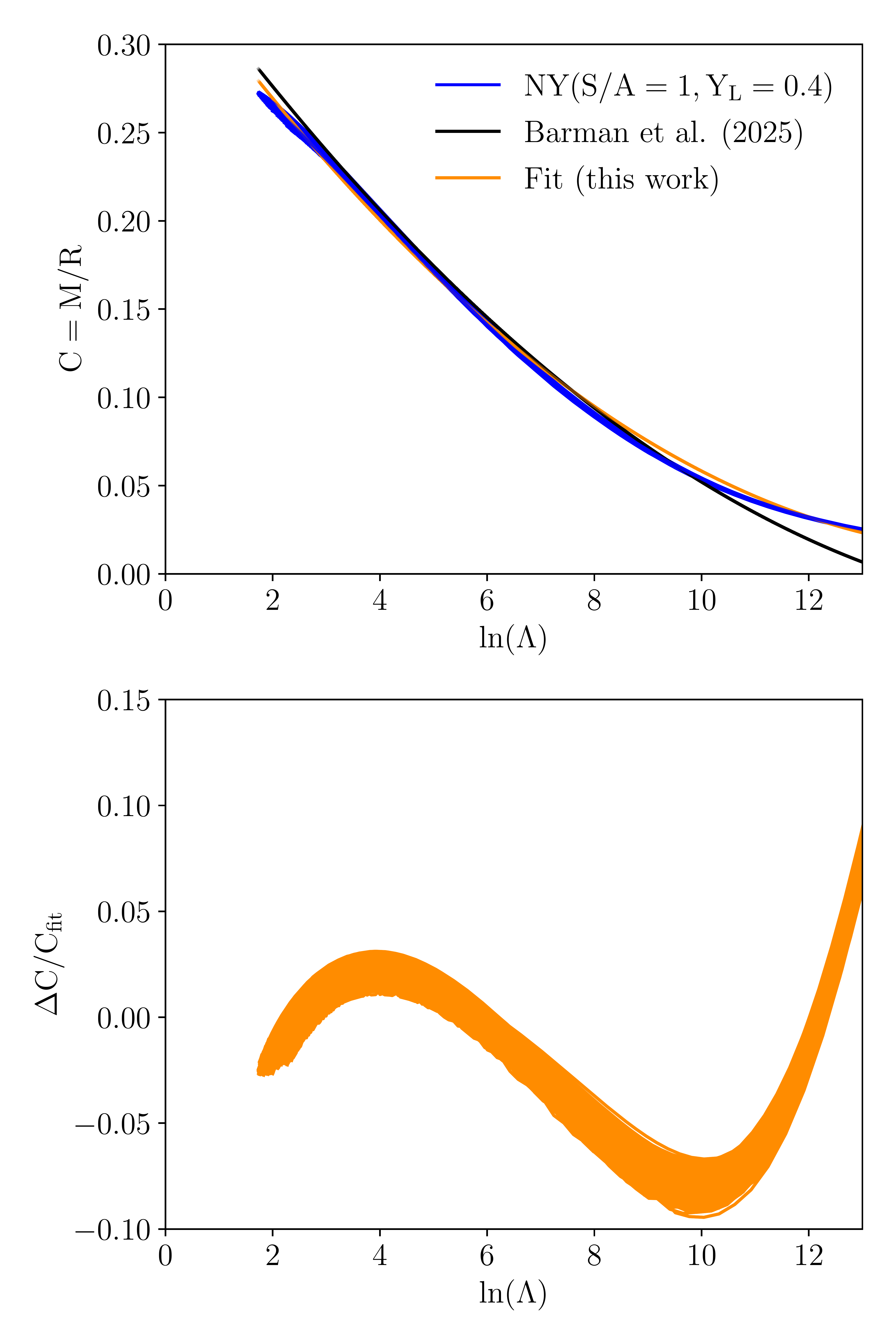}
        \caption{}
        \label{subfig:NY_1_0.4_C_Love}
    \end{subfigure}
    \begin{subfigure}{.45\linewidth}
        \centering
        \includegraphics[width=\textwidth]{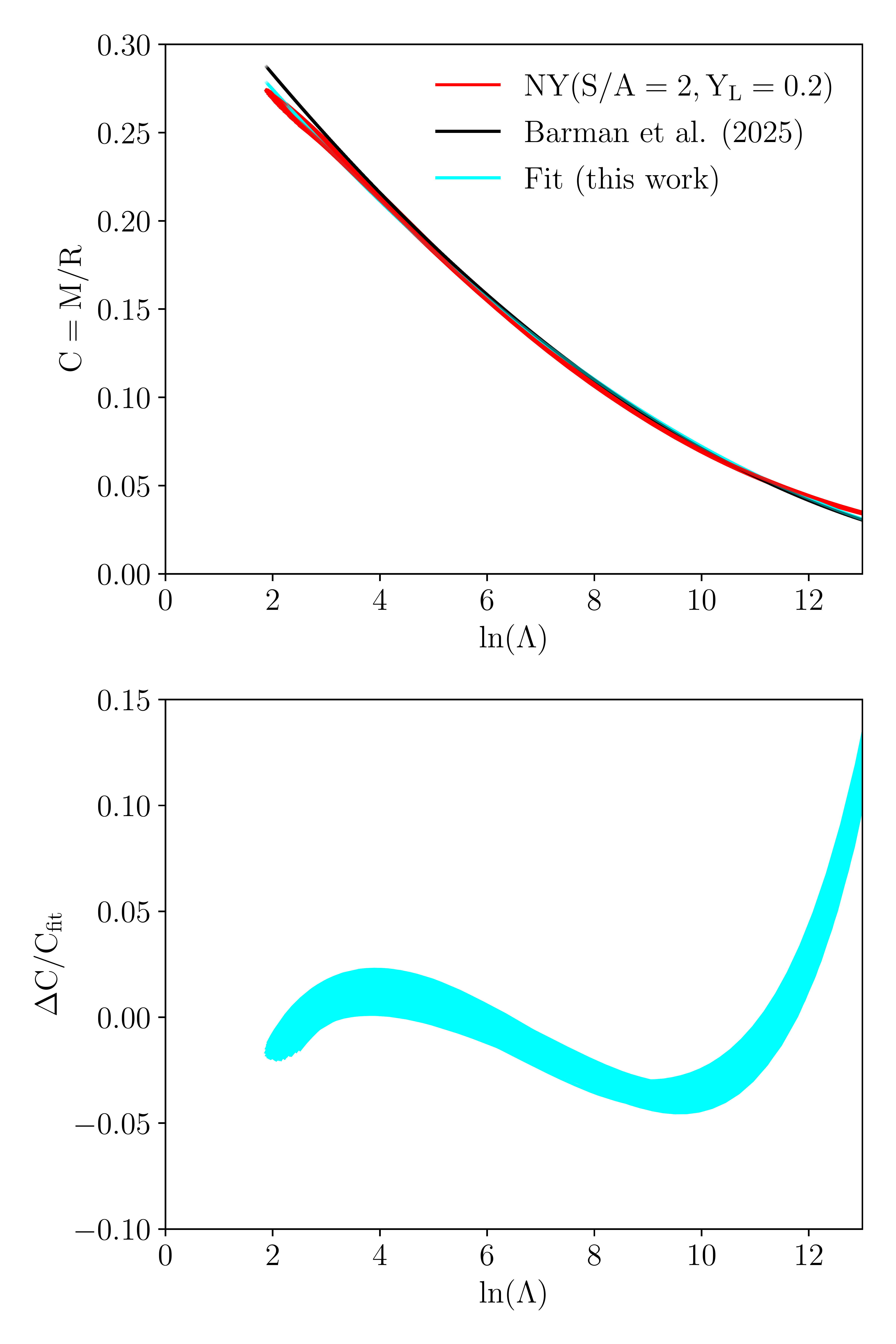}
        \caption{}
        \label{subfig:NY_2_0.2_C_Love}
    \end{subfigure}
    \caption{$C-Love$ relations. Top panel shows the universal relation and bottom panel shows the fit error obtained from fit relation(s) given in this work. The black curves in both rows represent the fit relation obtained in our previous work~\citep{Barman_Pradhan_2025prd}. The dark-orange/cyan curves represent fit relations found in this work. (a) $S/A=1$, $Y_L=0.4$; (b) $S/A=2$, $Y_L=0.2$}
    \label{fig:NY_C_Love}
\end{figure*}

We show our results in Fig.\ref{fig:NY_C_Love} and also compare with fit relations obtained in our previous study~\citep{Barman_Pradhan_2025prd} (black curves). We see that for both cases the universality is maintained subject to the constraints from $\chi$EFT + Astro. For the $NY(1,0.4)$ case (see Fig.~\ref{subfig:NY_1_0.4_C_Love}), we see that the universal relation deviates considerably from~\cite{Barman_Pradhan_2025prd} fit for higher tidal deformability ($\Lambda>20000$). These deviations are qualitatively less prominent for the other NS configuration (see Fig.~\ref{subfig:NY_2_0.2_C_Love}). \\

In~\cite{Barman_Pradhan_2025prd} fits were done for hot NS in the $\nu$-free domain and since we saw in Sec.~\ref{subsec:nu_free_vs_trapped} that neutrino trapping can slightly alter radii of the NS configurations, it is expected that the $C-Love$ relation would change due to neutrinos being present in the thermal bath. Qualitatively relative $\nu_e$ fractions are higher in the $NY(1,0.4)$ case and therefore the deviation from~\cite{Barman_Pradhan_2025prd} fit is also large. We provide our modified fit relations for both of these NS configurations in Table~\ref{tab:c_love_fit}. In the bottom panel of Fig.~\ref{fig:NY_C_Love}, we show the percentage error for these fit relations and found that they are less than 15\%.

\begin{table*}
    \centering
    \begin{tabular}{lccc}
        \toprule
        \toprule
         &$a_0$ & $a_1$ & $a_2$ \\
        \midrule
        $NY(1,0.4)$ & $3.490197\times10^{-1}$ & $-4.251316\times10^{-2}$ & $1.342965\times10^{-3}$ \\
        $NY(2,0.2)$ & $3.460195\times10^{-1}$ & $-3.788775\times10^{-2}$ & $1.049641\times10^{-3}$ \\
        \bottomrule
        \bottomrule
    \end{tabular}
    \caption{Fit relations for $C-Love$ relations found in this work for the two NS configurations. These relations are obtained after taking into account the uncertainty from $\chi$EFT and Astro.}
    \label{tab:c_love_fit}
\end{table*}

\section{Discussions} \label{sec:discussions}

{\it Summary:} In this work, we investigated $f$-mode oscillations in hot neutron stars relevant for astrophysical scenarios such as PNS and BNS mergers. For this, we developed an EOS framework by extending the non-linear Relativistic Mean Field Model to incorporate effects of finite temperature, hyperons and neutrino trapping consistently. This framework not only allows us to study the sensitivity to thermal conditions but also to nuclear and hypernuclear parameters, by systematically varying them within their allowed uncertainties to construct different nuclear parametrizations. In this way we were able to generate a large number of finite temperature EOSs compatible with multi-disciplinary physics inputs of $\chi$EFT, Astro and HIC and use them to infer the impact of nuclear/hypernuclear parameters on macroscopic properties and $f$-mode oscillations. \\

We first investigated the sensitivity of the model to thermal conditions for nucleonic (N) and hyperonic (NY) matter for fixed values of nuclear and hypernuclear parameters. We considered two isothermal configurations in the $\nu$-free and out-of-$\beta$-equilibrium state and found that due to thermal excitations at high temperature hyperonic species appear at lower densities. As the appearance of a hyperonic species softens the EOS, we see dips/kinks in $p_{th}$ and $\Gamma_{th}$ at density where such species appear for the cold case. This behavior was also seen for different EOS models in previous studies~\citep{Raduta_2022epja,Kochankovski_Ramos_2022mnras,Blacker_Kochankovski_2024prd}. \\

We then compared the sensitivity to thermal conditions for isentropic configurations and observed that neutrino-trapping had marginal impact on the macroscopic configurations for N-matter. For higher neutrino fractions, changes in radii were more significant for intermediate mass NS configurations. Additionally for the NY-matter, maximum mass supported by a thermal configuration with trapped neutrinos is somewhat less compared to the corresponding $\nu$-free case. There was a reversal in the stiffness of the thermal EOSs for N- and NY-matter. 
For both compositions, temperature as a function of density is suppressed in $\nu$-trapped regime due to the presence of an additional species. \\

To demonstrate the impact of various constraints on hot NS properties with NY-matter, we chose two neutrino trapped thermal configurations: $NY(1, 0.4)$ and $NY(2, 0.2)$. These thermal states represent respectively an early leptonized state and a late deleptonized state in the evolution of a PNS. We found that nuclear saturation density strongly influences properties of intermediate mass NS configurations for both the thermal cases. The correlations of $n_{sat}$ with $R_{2M_{\odot}}$, $\Lambda_{2M_{\odot}}$ and $\nu_{f,2M_{\odot}}$ significantly increased as a result of additional HIC constraints. The nucleon effective mass controls the stiffness of the EOSs and its effect weakens with the addition of HIC constraints. We also observed qualitatively similar correlations for both the thermal configurations. The other nuclear and hypernuclear parameters did not affect any observable properties (see appendices~\mbox{\ref{appendix_a} and~\ref{appendix_b}}).
\\

We note that the oscillation modes are calculated in the Cowling classification scheme. It has been shown that for the evolutionary phase before 0.4 seconds, these scheme may not hold~\citep{Rodriguez2023}. However, after this initial phase, Cowling scheme agrees well with different approaches. \\

{\it Comparison with other works:}
In our previous study in N-matter~\citep{Barman_Pradhan_2025prd}, we found that $m^*/m$ correlates with observable properties of high mass configurations as well as a change of correlation strengths between the two thermal configurations. Here, we did not see any such effects. The presence of hyperons can be attributed to this decreased effect of nucleon effective mass. We instead see that nuclear saturation density playing a more prominent role in determining structures of the intermediate mass hot NS configurations. The reversal of stiffness for different thermal configurations for N- and NY-matter was also previously shown in the paper by~\cite{Thapa_Beznogov_2023prd}, where a selection of $\nu$-free tabulated EOSs were used. In the same paper, the authors showed that the appearance of hyperons or other species in the hot NS lead to an increase in $f$-mode frequency compared to that of the N-matter star for the same gravitational mass. \\

\cite{Tsiopelas_Sedrakian_2024} in their recent work have systematically varied $L_{sym}$ as 30, 50 and 70 MeV and generated EOSs for N- and NY-matter. In that work, the authors found that the effect of slope of symmetry energy on the EOS diminishes with increasing temperature. In our previous work with cold matter~\citep{Pradhan_Chatterjee_2022prc}, it was shown that the correlation between $L_{sym}$ and astrophysical observables does not appear in all nuclear models, and it depends on which nuclear parameters control the high density behavior of the EOS. Further,~\cite{Jaiswal2021} found within the NL-RMF model that effective mass (and not slope of symmetry energy) to be strongly correlated with astrophysical observables for N-matter. It is therefore important to systematically vary all the nuclear and hypernuclear parameters within a Bayesian scheme to investigate which nuclear parameters show physical correlations among themselves and with NS observables, as demonstrated in our work. \\

We also investigated the universal relation between $f$-mode frequency and average mass density in case of two neutrino trapped hot NY-matter configurations and compared them with the previous fit relations given for cold and $\beta$ stable matter by various authors. We found that~\cite{Andersson_Kokkotas_1998mnras} and~\cite{Pradhan_Chatterjee_2021prc} fits underestimate the $f$-mode frequencies for both cases while~\cite{Doneva_Gaertig_2013prd} fit overestimates the same. We provide the modified fit relations for both the cases constrained by $\chi$EFT + Astro.
We also calculated the $C-Love$ relations for these hot NS configurations and found that nuclear and hypernuclear parameters do not affect these relations. However, $NY(1,0.4)$ state which supports higher degree of neutrino trapping show more deviation from the universal relation without neutrino trapping. We provide our modified second order fit relations for these configurations. \\

We note that previous study by~\cite{Li_Sedrakian_2023apj} consistently vary $L_{sym}$ and higher order parameter $Q_{sat}$ using RMF model with a Density-Dependent (DD) coupling formalism. Bayesian approaches with cold matter have been employed using DD RMF models in previous studies~\citep{Li_Tian_2025prc,Beznogov_Raduta_2023prc,Li_Sedrakian_2025prc,Li_Tian_2025plb}. These DD based Bayesian studies can be extended to finite temperature to explore impact of nuclear and/or hypernuclear properties on neutrino trapped hot NS configurations and their oscillation modes. Similarly this formalism with non-linear RMF model involving additional non-linear couplings such as vector self-interaction can be employed to this Bayesian analysis.
\\

Our study hopes to motivate the importance of taking into account the still persistent uncertainty in the nucleonic and hyperonic sectors while constructing hot EOS for studying Proto-Neutron Star configurations as well as to relativistic simulations of CCSN and BNS mergers. \cite{Kochankovski_Ramos_2024mnras} showed the importance of hypernuclear potentials in the EOS of hot and dense NY-matter by taking the upper and lower limits of these potentials. The nuclear parameters in that study were kept fixed. We showed that by varying these parameters and taking large number of EOSs compatible with different constraints, the nuclear saturation properties play a dominant role over other nuclear and hypernuclear uncertainties. 
We provide tables for the finite temperature EOSs compatible with current constraints in the CompOSE database for possible application in numerical simulations.

\section*{Acknowledgments}
N.B. and D.C. thank L. Suleiman and M. Oertel for providing valuable insight regarding $\nu$-trapping in hot and dense matter. The authors are also grateful to the fruitful discussion with S. Ghosh and B. K. Pradhan about imposing different constraints mentioned in this paper.

\section*{Data Availability}
No new data was generated in support of this document. The data used in this work are available from the corresponding author upon reasonable request.

\bibliographystyle{mnras}
\bibliography{thebibliography}

\appendix 
\section{Correlation among nuclear and hypernuclear parameters with observables for constraints from $\chi$EFT + Astro} \label{appendix_a}

\begin{figure}
    \centering
    \includegraphics[width=1.05\linewidth]{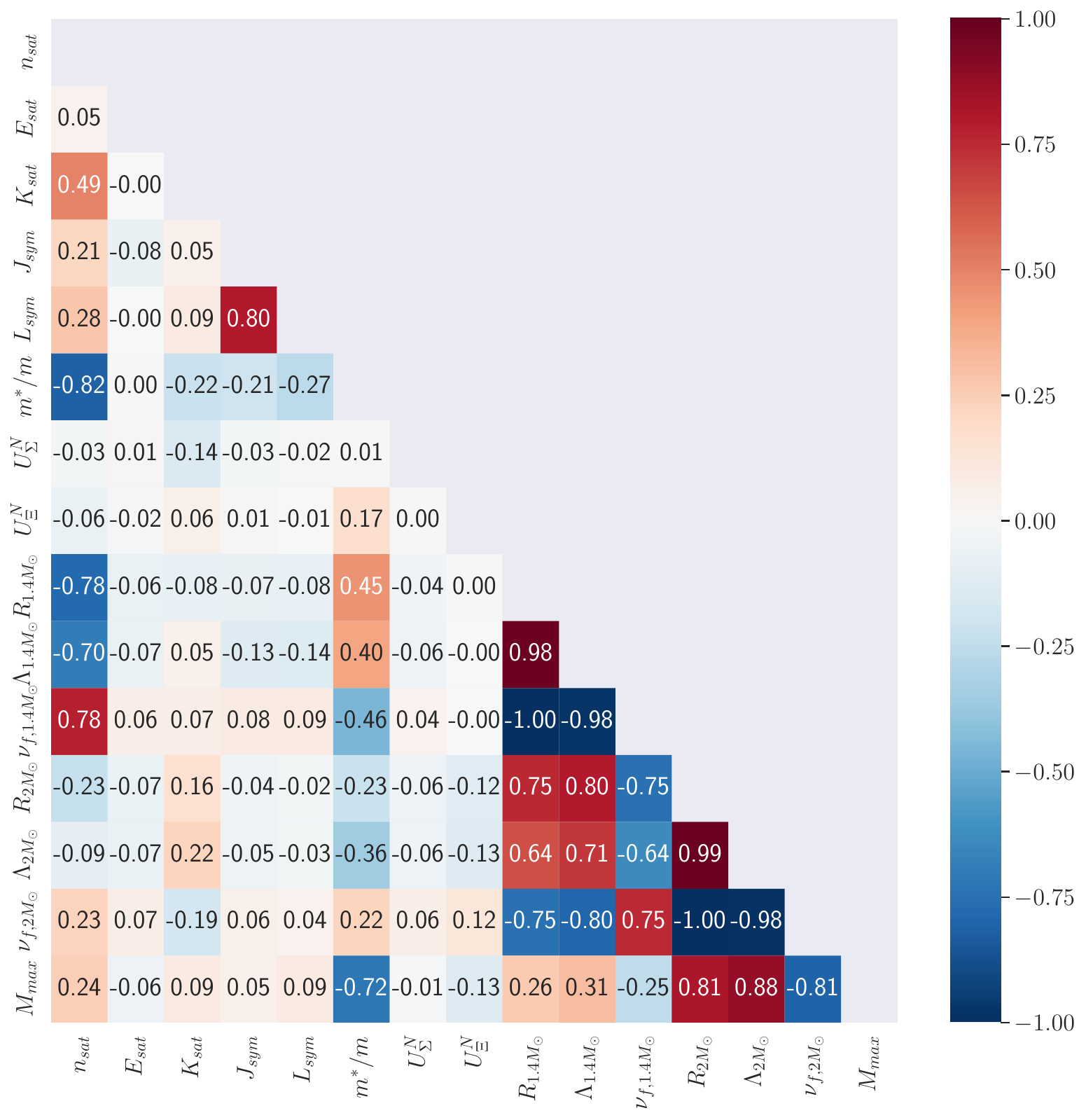}
    \caption{Correlation matrix for $NY(1,0.4)$ configuration found from constraints from $\chi$EFT + Astro}
    \label{fig:correlation_NY_1_0.4_nu}
\end{figure}

We show the complete correlation matrix that describes how nuclear and hypernuclear parameters affect observables of the $NY(1, 0.4)$ NS configuration in Fig.~\ref{fig:correlation_NY_1_0.4_nu} with constraints from $\chi$EFT + Astro. Among nuclear saturation properties, $n_{sat}$ shows strong anti-correlation with $m^*/m$ and there exists a positive correlation between asymmetry parameters $J_{sym}$ and $L_{sym}$. $n_{sat}$ is strongly correlated with $R_{1.4M_{\odot}}$, $\Lambda_{1.4M_{\odot}}$ and $\nu_{f,1.4M_{\odot}}$. The nucleon effective mass is found to be anti-correlated with $M_{max}$. No correlation of hypernuclear parameters are observed with any nuclear saturation property and observable. We note that correlation matrix for $NY(2,0.2)$ configuration with the same constraints is not shown as the result is found to be qualitatively similar. We see that hypernuclear potentials ($U^N_{\Sigma}$ and $U^N_{\Xi}$) are seen to not be correlating with any other property.

\begin{figure}
    \centering
    \includegraphics[width=1.05\linewidth]{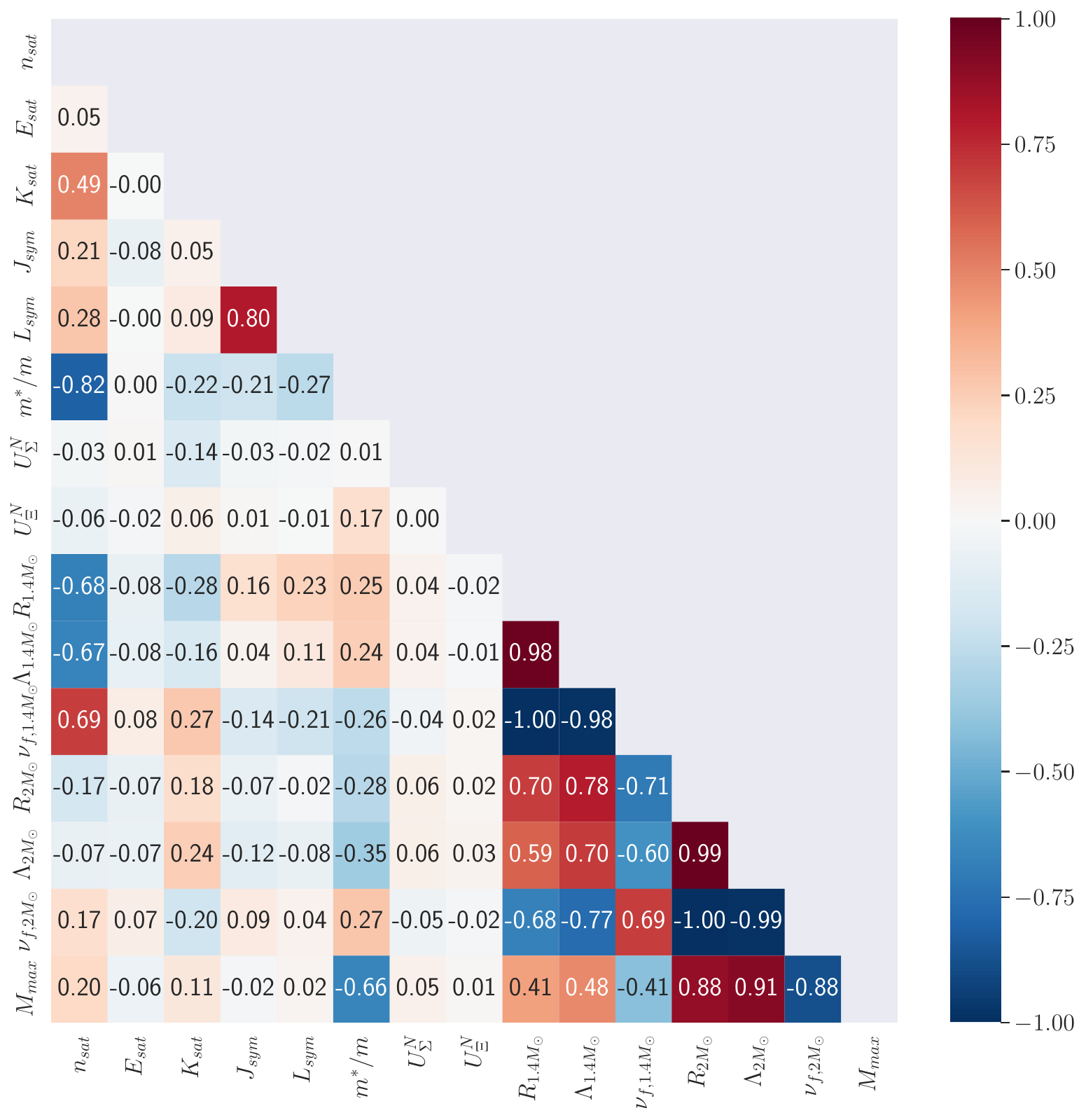}
    \caption{Same as~\ref{fig:correlation_NY_1_0.4_nu} but for $NY(2,0.2)$ configuration}
    \label{fig:correlation_NY_2_0.2_nu}
\end{figure}

\section{Correlation among nuclear and hypernuclear parameters with observables for constraints from $\chi$EFT + Astro + HIC} \label{appendix_b}

\begin{figure}
    \centering
    \includegraphics[width=1.05\linewidth]{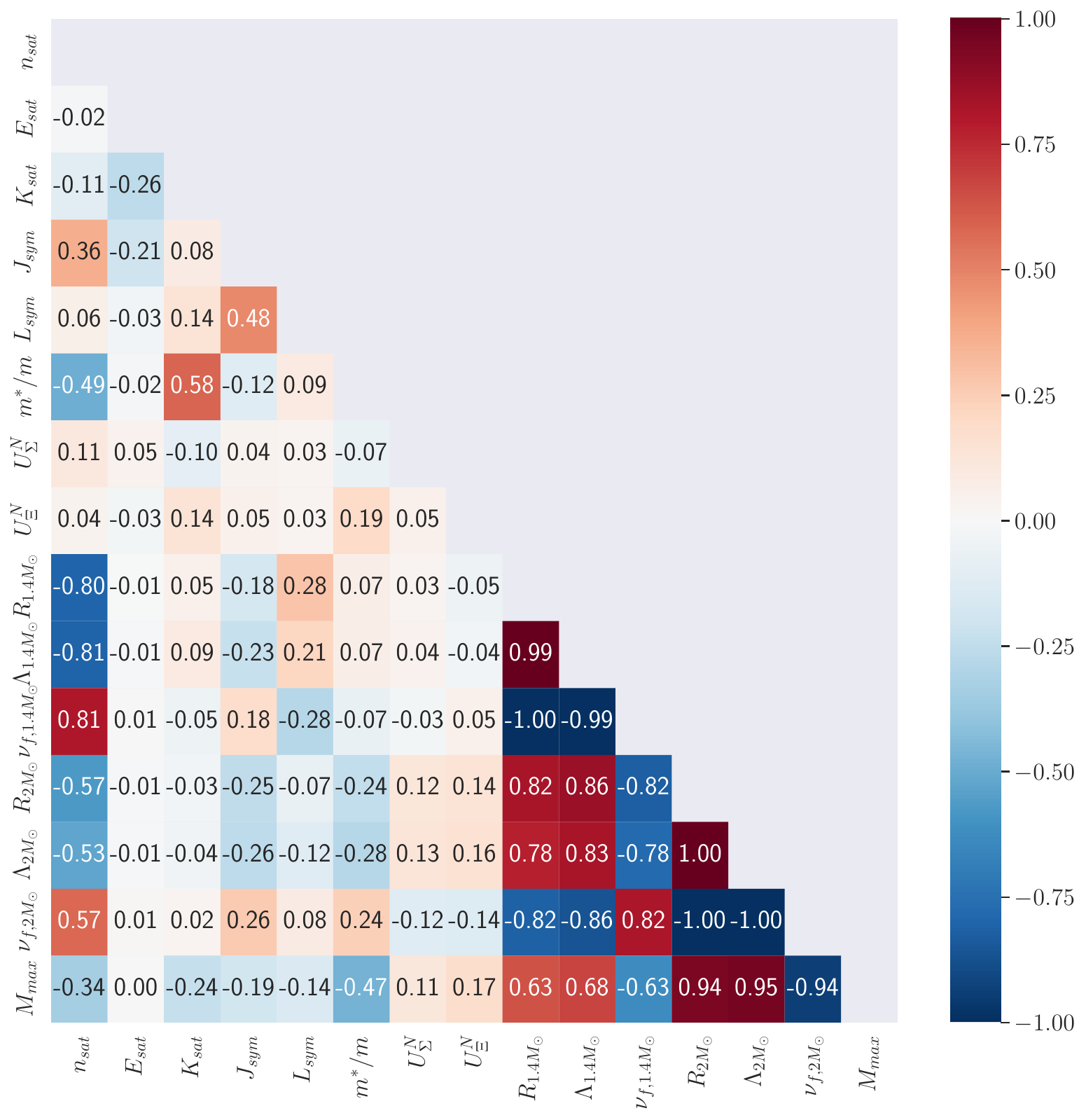}
    \caption{Correlation matrix for $NY(2,0.2)$ configuration found from constraints from $\chi$EFT + Astro + HIC}
    \label{fig:correlation_NY_2_0.2_nu_hic}
\end{figure}

We show the impact of nuclear and hypernuclear parameters on the observables of the $NY(2, 0.2)$ configuration in Fig.~\ref{fig:correlation_NY_2_0.2_nu_hic}. There is noticeable increase in the correlation/anti-correlation of $n_{sat}$ with observables of canonical $2M_{\odot}$ hot NS configuration. On the other hand, the correlation of $m^*/m$ with $M_{max}$ has been drastically reduced. Another observation is the increase in the correlation between the effective mass and the compressibility parameter, $K_{sat}$. Again we observe that hypernuclear properties are not affecting any other parameters or observables.

\end{document}